\documentclass[12pt]{article}

\usepackage{latexsym,amsmath,amssymb,theorem,epsfig}

\usepackage{graphicx}
\usepackage{graphicx,subfigure}
\usepackage{epstopdf}
\usepackage[body={17.5cm, 21cm},right=2cm]{geometry}
\usepackage{amssymb}
\usepackage{amsmath}
\usepackage{psfrag}
\usepackage{epsfig}
\usepackage{cancel}
 \allowdisplaybreaks[4]

\usepackage[all]{xy}

\DeclareFontFamily{OT1}{rsfs10}{}
\DeclareFontShape{OT1}{rsfs10}{m}{n}{ <-> rsfs10 }{}
\DeclareMathAlphabet{\mathscript}{OT1}{rsfs10}{m}{n}

\numberwithin{equation}{section}

\newcommand{\ns}{\normalsize}

\newcommand{\be}{\beta}

\newcommand{\kk}{{\bf k}}

\newcommand{\comb}{(1-\epsilon-\epsilon_s)}
\newcommand{\combtwo}{(1+\epsilon)}

\newcommand{\es}{\epsilon_s}

\def\gsim{ \lower .75ex \hbox{$\sim$} \llap{\raise .27ex \hbox{$>$}} }
\def\lsim{ \lower .75ex \hbox{$\sim$} \llap{\raise .27ex \hbox{$<$}} }
\def\be{\begin{equation}}
\def\ee{\end{equation}}
\def\bea{\begin{eqnarray}}
\def\eea{\end{eqnarray}}

\begin{document}

\begin{titlepage}

\vspace{-5cm}

\title{
  \hfill{\ns }  \\[1em]
   {\LARGE Rapidly-Varying Speed of Sound, Scale Invariance \\ and Non-Gaussian Signatures}
\\[1em] }
\author{
   Justin Khoury and Federico Piazza
     \\[0.5em]
   {\ns Perimeter Institute for Theoretical Physics} \\[-0.2cm]
{\ns Waterloo, Ontario, N2L 2Y5, Canada}\\[0.3cm]}

\date{}

\maketitle

\begin{abstract}
We show that curvature perturbations acquire a scale invariant spectrum for any constant equation of state, provided the
fluid has a suitably time-dependent sound speed. In order for modes to exit the physical horizon, and in order
to solve the usual problems of standard big bang cosmology, we argue that the only allowed possibilities are inflationary
(albeit not necessarily slow-roll) expansion or ekpyrotic contraction. Non-Gaussianities offer many distinguish features.
As usual with a small sound speed, non-Gaussianity can be relatively large, around current sensitivity levels.
For DBI-like lagrangians, the amplitude is negative in the inflationary branch, and can be either negative or positive in the ekpyrotic branch.
Unlike the power spectrum, the three-point amplitude displays a large tilt that, in the expanding case, peaks on smallest  scales. While the shape is predominantly of the equilateral type in the inflationary branch, as in DBI inflation, it is of the local form in the ekpyrotic branch. The tensor spectrum is also generically far from scale invariant.
In the contracting case, for instance, tensors are strongly blue tilted, resulting in an unmeasurably small gravity wave amplitude on
cosmic microwave background scales.
\end{abstract}

\thispagestyle{empty}

\end{titlepage}

\section{Introduction}

The observation that the large scale structure originated from a nearly scale invariant primordial spectrum is generally construed as evidence that the universe underwent a phase of quasi-de Sitter expansion. Given the limited number of early universe parameters constrained by observations thus far, however, it is prudent to keep an open mind and seek the most general mechanism
for producing density perturbations consistent with the data. It is well-known that generating a scale invariant spectrum of curvature perturbations with a single scalar field in 3+1 dimensions requires either inflation or a contracting dust-dominated phase~\cite{dust,gratton}. 
On the other hand, scale invariant entropy perturbations can be generated in a larger variety of cosmological scenarios, provided more than one scalar field is present. 
This has been shown to occur, for example,  in the slowly contracting background of the new ekpyrotic cosmology~\cite{newek,fabio,lehners,paolo,wands,notari} or in the kinetic dominated pre-big bang cosmology~\cite{pbb,pump}. (See~\cite{jlrev} and~\cite{pbbrev} for recent reviews of the ekpyrotic/cyclic and pre-big bang scenarios, respectively.) Entropy perturbations can then be converted into curvature perturbations, for instance if the curvaton mechanism \cite{mollerach, curvaton, pbbcurvaton} is at work. These earlier studies all have in common a constant sound speed $c_s$ for scalar field fluctuations. 

In this paper we show that allowing for a time-dependent sound speed greatly broadens  the range of possible cosmologies. The curvature perturbation on uniform density hypersurfaces $\zeta$ can acquire a scale invariant spectrum for arbitrary background equation of state parameter $\epsilon = -\dot{H}/H^2 = 3(1+w)/2$, as long as the sound speed varies appropriately.   

Although our derivation applies to any perfect fluid, we take as our fiducial model a single scalar field $\phi$ with a Lagrangian 
\begin{equation} \label{lag}
{\cal L}_\phi = \sqrt{-g} P(X,\phi)
\end{equation} 
which is a general function of the kinetic term 
$X = -\frac{1}{2} g^{\mu \nu}\partial_\mu \phi \partial_\nu \phi$ and the field itself. In this theory the perturbations around a homogeneous background are still adiabatic. However, the presence of a non-standard kinetic term allows for a background dependent ``speed of sound" of the fluctuations\footnote{Analogously, higher order
curvature terms can induce a sound speed of tensor perturbations~\cite{maurizio}.} 
\begin{equation}
c_s^2 = \frac{P_{,X}}{\rho_{,X}}= \frac{P_{,X}}{P_{,X}+2X P_{,XX}}\, ,
\end{equation}
where $\rho = 2 X P_{,X} - P$ is the energy density of the field, and the quantities above are calculated on the unperturbed solution. While our basic mechanism for generating a scale invariant spectrum only relies on a varying sound speed of the adiabatic perturbations, we will refer more specifically to the single scalar field model \eqref{lag} for calculating non-Gaussianities. 

In terms of the rate of change of the sound speed, $\epsilon_s \equiv \dot{c_s}/c_sH$, exact scale invariance requires $\epsilon$, $\epsilon_s$ to be constant and to satisfy a relation between them. More precisely, we find two branches of solutions: 

\begin{itemize}

\item Case I: Expanding universe with decreasing sound speed, such that $\epsilon_s = -2\epsilon$.

\item Case II: Contracting universe with increasing sound speed, such that $\epsilon_s = \frac{2}{5}(3-2\epsilon)$.

\end{itemize}

In the expanding case this mechanism was first proposed by~\cite{cri} (see also \cite{joao,piao}), whereas the contracting branch of solutions is new.
Unlike~\cite{joao}, who considered a universe with superluminal sound speed initially, here we restrict our analysis to subluminal sound speeds throughout. (Theories with superluminal propagation have been argued not have a consistent UV-completion~\cite{IRUV}.)

In the special case $\epsilon_s = 0$, these recover de Sitter expansion and dust-like contraction, respectively. Of course, small deviations from these conditions and/or mild time-dependence for $\epsilon$,$\epsilon_s$ lead to small departures from scale invariance. In order for perturbations to be stretched outside the Hubble horizon, the equation of state must satisfy $\epsilon < 1$ in Case I, corresponding to inflationary expansion, or $\epsilon > 1$ in Case II, corresponding to decelerated contraction. While the expanding branch requires inflation, it is worth emphasizing that our mechanism is not constrained by slow-roll conditions and therefore encompasses a much broader range of inflationary cosmologies.

The above cosmologies, while degenerate at the level of the power spectrum, can be distinguished through their non-Gaussian signatures.  To begin with, the sign of the three-point amplitude is generically negative in Case I and can be either negative or positive in Case II. More importantly, unlike standard inflationary models, here the near scale invariance of the two-point function does not extend to the three-point amplitude. The non-Gaussian amplitude is proportional to $1/c_s^2$ evaluated at horizon crossing, and therefore exhibits strong running with scale~\cite{runningchen,loverde}. In particular, non-Gaussianity peaks on the smallest scales in Case I since $c_s$ decreases in time, whereas the opposite is generally true in Case II. The sign of the tilt therefore allows to distinguish between the expanding and contracting branches.

The shape of the three-point function in momentum space is also characteristically different in the two branches. In the expanding case, the three-point amplitude peaks
for {\it equilateral} momentum triangles~\cite{paoloshape}, as usual for models with small sound speed such as Dirac-Born-Infeld (DBI) inflation~\cite{eva,DBI2,chen}. In the contracting case, however, the non-Gaussian signal features a strong peak for squeezed triangles, corresponding to the so-called {\it local} form~\cite{komatsuspergel}. This may come as a surprise since local non-Gaussianities usually require two scalar fields, as in the curvaton~\cite{curvaton}, modulated reheating~\cite{DGZ}, pre-big bang pump field~\cite{pump} and New Ekpyrotic mechanisms~\cite{paolo,ngek,ngekwands,ngcyclic}. 

In fact, the striking difference in shape between Cases I and II traces back to the evolution of $\zeta$ at long wavelengths. In the expanding branch, the growing mode of $\zeta$ is a constant, and non-linearities have therefore a finite time to grow until modes go outside the horizon. Maximal amplification is achieved when all wavevectors have magnitude comparable to the Hubble radius,
leading to the predominantly equilateral shape. In the contracting branch, on the other hand, $\zeta$ keeps growing on super-Hubble scales, thereby allowing non-linearities to grow
well after modes have exited the horizon. The non-Gaussian amplitude therefore peaks when one of the wavenumbers is small, corresponding to the squeezed-triangle limit.

The behavior of $\zeta$ on super-Hubble scales is closely related to the issue of stability of the background. In the long-wavelength limit, $\zeta$ represents a mildly scale-dependent perturbation in the scale factor~\cite{bondsalopek,malda}. Thus all Case I cosmologies are stable, since a constant $\zeta$ can be absorbed in a spatial diffeomorphism. Meanwhile, the super-Hubble growth of $\zeta$ in Case II indicates that these backgrounds are unstable. Remarkably, such tachyonic instabilities appear to be a ubiquitous feature of contracting mechanisms: both New Ekpyrotic Cosmology and the general class of contracting two-field models of~\cite{tolleywesley} display a tachyonic instability in the entropy direction. This by no means represents a show-stopper --- it was shown in~\cite{newekinit}, for instance, that a pre-ekpyrotic stabilizing phase can bring the field trajectory arbitrarily close to the desired classical solutions for a broad range of initial conditions. Moreover, as the present work underscores, the super-horizon growth of $\zeta$ leads to striking features in non-Gaussianity.

Another important observable are primordial gravitational waves. Since tensor modes are oblivious to the time dependence of the sound speed and only feel the background cosmology, their spectrum is in general not scale invariant. In the expanding case, the gravity wave amplitude peaks on large scales, corresponding to a red tensor spectral tilt. In the contracting case, gravity waves  are peaked on small scales, corresponding to a strong blue tilt.

Various phenomenological considerations constrain the allowed equation of state parameter. We have already mentioned the inflationary condition $\epsilon < 1$ in Case I, in order for perturbations to exit the Hubble horizon. However, this range of $\epsilon$ of course also ensures that (moderate amounts of) spatial curvature and anisotropy become increasingly irrelevant
in time. If we impose similar standards in Case II, we obtain $\epsilon > 3$ (or $w>1$), corresponding to the ekpyrotic range. Indeed, ekpyrotic contraction is equally powerful as inflation in
solving the homogeneity, isotropy and flatness problems of standard big bang cosmology~\cite{gratton,mixmaster}. Furthermore, the tilt of non-Gaussianities cannot be too strong, for otherwise perturbations risk exiting the perturbative regime within the observable range (Case I) or lead to an unacceptably large amplitude on cosmic microwave background (CMB) scales (Case II). These non-Gaussian considerations lead to $\epsilon \lesssim 0.5$  in Case I. In Case II, on the other hand, non-Gaussianities are generically much larger and constrain the model in the region of the $c_s - \epsilon$ parameter space plotted in Fig.~\ref{fit1}. Finally, since the tensor spectrum is red in the expanding case, we must ensure that the gravitational wave amplitude on large scales is consistent with CMB bounds. This requires the slighly tighter bound: $\epsilon \lesssim 0.3$ (Case I).

While our analysis applies to general scalar field models of the form~(\ref{lag}), we also provide explicit microphysical realizations of the scale invariant conditions within the context of DBI effective lagrangians. Here we piggyback on~\cite{kinney}, where it was shown that any scaling solution for the scale factor and sound speed can be obtained by suitably choosing the warp factor $f(\phi)$ and scalar potential $V(\phi)$. Remarkably, the desired condition $\epsilon_s = -2\epsilon$ is satisfied (at least in the regime of small sound speed) for the AdS warp factor and $m^2\phi^2$ potential originally studied by~\cite{eva}. Thus scale invariance obtains in this case {\it without} requiring the slow roll condition $\epsilon \ll 1$.

To conclude this introduction, we offer an executive summary of our key results:

\begin{itemize}

\item A scale invariant spectrum of perturbations can be obtained for any background cosmology with constant equation of state, by suitably choosing the time dependence of the sound speed. We find both an expanding (Case~I) and a contracting (Case II) branch of solutions. We discuss this basic mechanism in Secs.~\ref{2} and~\ref{ampstab}.

\item In the expanding case, various phenomenological bounds constrain the equation of state parameter to $\epsilon \lesssim 0.3$  and are discussed in Sec.~\ref{4}. Our inflationary range is much broader than allowed by the usual slow-roll conditions.

\item Scale invariance requires $\epsilon$,$\epsilon_s$ to be constant and to satisfy a specific relation (see above). Deviations from these conditions lead to a non-zero spectral tilt, which is derived in Sec.~\ref{ns}.

\item Gravitational waves are far from scale invariant. The tensor spectral index, derived in Sec.~\ref{gw}, is negative in Case I and positive in Case II, corresponding to red and blue tilts, respectively.

\item Relatively large non-Gaussianities are generated in both cases (Secs.~\ref{7} and~\ref{ngsign}). In Case I they are of the DBI-equilateral type and generically negative; in Case II they are of the local type and generically positive. 

\item A (nearly) scale invariant two-point function does not necessarily imply a (nearly) scale invariant three point function. In fact, our mechanism fails to protect against  running of non-Gaussianities. In Case I the three-point function has a strong blue tilt; in Case II it generally has a strong red tilt (see Sec.~\ref{running}).

\end{itemize}

\section{Scale Invariance from Time-Dependent Sound Speed} \label{2}

Our key observation is that in the presence of a time-dependent sound speed $c_s$ the curvature perturbation evolves
according to an effective cosmological background different from the actual scale factor. 
Thus, a scale-invariant spectrum of density perturbations can arise for any background equation of state parameter 
$\epsilon = -\dot{H}/H^2$, for a suitably-varying $c_s$. 
In this Section, we show that scale invariance can also obtain in the contracting case as well.

At quadratic order, the action for the curvature perturbation $\zeta$ for general speed of sound models is given by~\cite{garrigamukhanov}
\be \label{actionzeta}
S = \frac{M_{\rm Pl}^2}{2}\int {\rm d}^3x{\rm d}\tau \;z^2\left[ \left(\frac{d\zeta}{d\tau}\right)^2 - c_s^2(\vec{\nabla}\zeta)^2\right]\,,
\ee
where $\tau$ is conformal time, and $z$ is defined as usual by $z = a \sqrt{2 \epsilon}/c_s$.
This holds whether the matter is a perfect fluid or a $P(X,\phi)$ scalar field.
Since $c_s$ is a general function of time, it is convenient to instead work in terms of
the ``sound-horizon" time ${\rm d}y = c_s {\rm d}\tau$:
\be
S = \frac{M_{\rm Pl}^2}{2}\int {\rm d}^3x{\rm d}y \;q^2\left[  \zeta'^2 -(\vec{\nabla}\zeta)^2\right]\,,
\ee
where $'\equiv {\rm d}/{\rm d}y$, and
\be
q \equiv \sqrt{c_s}z = \frac{a \sqrt{2 \epsilon}}{\sqrt{c_s}}\,.
\label{qdef}
\ee
The virtue of this time redefinition is manifest: the kinetic term takes the standard form, as if $c_s = 1$,
while the dependence on $c_s$ has been absorbed in a modified background
described by $q$.

In terms of the canonically-normalized scalar variable $v = M_{\rm Pl} q\zeta$, the equations of motion for the Fourier modes
are given by
\be
v''_k +\left(k^2 - \frac{q''}{q}\right)v_k = 0\,.
\label{veqn}
\ee
It is well-known that this results in a scale-invariant spectrum if and only if $q''/q = 2/y^2$. Indeed,
for the adiabatic vacuum choice, the mode functions are given by
\be 
v_k(y) = \frac{1}{\sqrt{2k}}\left(1-\frac{i}{ky}\right)e^{-iky}\,,
\label{vsol}
\ee
resulting in a scale-invariant $P_v(k) \sim k^3v_k^2$ in the limit $k\rightarrow 0$. Therefore, there are
two possibilities:
\be
q \sim (-y)^{-1}  \qquad {\rm or}\qquad q \sim y^2 \,,
\label{qsi}
\ee
where $y$ runs from $-\infty$ to $0$. In the limit of constant $c_s$, these of course reduce to
de Sitter expansion and dust-dominated contraction~\cite{dust,gratton}, respectively. But, as we now show,
more general backgrounds can also do the trick for suitably-chosen $c_s$.

To proceed, let us assume for simplicity that the background has constant equation of state: $\epsilon = {\rm const.}$ Similarly, as usual
we can define a rate of change parameter $\epsilon_s$ for the sound speed,
\be
\epsilon_s = \frac{\dot{c_s}}{Hc_s}\,,
\ee
also assumed to be constant. (In Sec.~\ref{ns}, we will relax these assumptions when calculating the spectral tilt.) From the definition of $y$,
it is then straightforward to show that
\be
a(y)\sim (-y)^{\frac{1}{\epsilon_s + \epsilon - 1}}\;;\qquad c_s(y) \sim (-y)^{\frac{\epsilon_s}{\epsilon_s + \epsilon - 1}}\,.
\label{const}
\ee
Substituting these into~(\ref{qdef}), then from~(\ref{qsi}) we deduce that scale-invariance occurs either for
\be
\epsilon_s = -2\epsilon \qquad  ({\rm Case\;I})\,,
\label{caseI}
\ee
corresponding to an {\it expanding background} with {\it decreasing sound speed}, or
\be
\epsilon_ s = \frac{2}{5}(3- 2 \epsilon)\qquad  ({\rm Case\;II})\,,
\label{caseII}
\ee
corresponding to a {\it contracting universe} with {\it growing sound speed} (for $\epsilon > 3/2$). Case I was
pointed out in~\cite{cri,joao,piao} (see also \cite{kinney} for a realization in DBI models of inflation~\cite{eva}), whereas Case II is new. As a check, with $\epsilon_s = 0$ this recovers
de Sitter expansion ($\epsilon = 0$) or dust-like contraction ($\epsilon = 3/2$). In slow-roll inflation, $\epsilon$ is by definition small, and
scale invariance forces $\epsilon_s$ to be small as well. What the above results show, however, is that much larger values of $\epsilon$ are
allowed as well, provided $\epsilon_s$ is chosen appropriately.

\section{Amplitude of Perturbations and Stability}
\label{ampstab}

From the explicit solution for $v_k$ in~(\ref{vsol}), we can read off the long-wavelength amplitude for $\zeta_k = v_k/q$:
\be
k^{3/2}\left\vert\zeta_k\right\vert = \frac{\sqrt{c_s}}{2M_{\rm Pl}\sqrt{\epsilon}ay}\,.
\ee
This can be further simpified by noting that~(\ref{const}) allows to solve for $y$ in terms $H$, $a$ and $c_s$. Substituting the result
into the above amplitude, the $\zeta$ power spectrum is readily obtained:
\be
P_\zeta \equiv \frac{1}{2\pi^2}k^3\left\vert\zeta_k\right\vert^2 = \frac{1}{8\pi^2\epsilon}(\epsilon_s + \epsilon-1)^2\frac{H^2}{c_sM_{\rm Pl}^2}\,.
\label{zetaPk}
\ee
In the quasi-de Sitter and nearly constant $c_s$ limit, $\epsilon \ll 1$ and $\epsilon_s\ll 1$,
this agrees with earlier results~\cite{garrigamukhanov}. 

Although the above expression for $P_\zeta$ applies both to Cases I and II, the time-dependence of the right-hand side is
different in each case. In Case I, $H^2/c_s$ is constant --- see~(\ref{const}) ---, and, following standard conventions,
it therefore makes sense to evaluate $P_\zeta$ at horizon-crossing (denoted by ``bars"):
\be 
P_\zeta^{({\rm I})}  = \frac{(1 + \epsilon)^2}{8\pi^2\epsilon}\frac{\bar{H}^2}{\bar{c}_sM_{\rm Pl}^2}\,,
\label{PI}
\ee
where we have substituted~(\ref{caseI}).

In Case II, on the other hand, $H^2/c_s\sim 1/y^6$, and the amplitude for $\zeta$ keeps growing outside the horizon. 
Moreover, unlike Case I, here the final amplitude depends not only on the dynamics during the scale invariant phase, but
also on the details of whatever phase precedes the bounce. For definiteness, we will assume that super-horizon modes stop growing
after the scale invariant phase. This is achieved, for instance, if $c_s$ saturates to a constant value and the collapse carries on
with an equation of state $w>1$, or $\epsilon >3$. In this case, the final amplitude of the power spectrum is evaluated at
the end of the scale-invariant phase:
\be
P_\zeta^{({\rm II})}  = \frac{(1 + \epsilon)^2}{200\pi^2\epsilon}\frac{H_{\rm end}^2}{c_{s\;{\rm end}}M_{\rm Pl}^2}\,,
\label{PII}
\ee
where now we have substituted~(\ref{caseII}). 

Of course this difference in time-dependence is intimately tied to the stability of the background. In the infinite-wavelength or $k\rightarrow 0$
limit, classical solutions for $\zeta$ are interpreted as homogeneous perturbations to the background scale factor~\cite{bondsalopek,malda}:
\be
ds^2 = a^2(\tau)(-{\rm d}\tau^2 + e^{2\zeta}{\rm d}\vec{x}^2)\,.
\ee
From $\zeta = v/q$ and the explicit solution~(\ref{vsol}), we see that the growing mode is $\zeta \sim 1/qy$.
In Case I, which includes standard inflation, $q\sim 1/y$, and $\zeta\rightarrow {\rm const.}$ on super-Hubble scales~\cite{bst,proof}. Thus these backgrounds are stable.
In Case II, however, $q\sim y^2$, and the growth of $\zeta$ as $k\rightarrow 0$ indicates that the background is not an attractor.

This instability appears to be a ubiquitous feature of all contracting mechanisms for generating perturbations.
In the two-field ekpyrotic mechanism, for instance, the entropy direction must be tachyonic~\cite{newek,wands,newekinit}.
This was shown to be true for arbitrary two-derivative action of two scalar fields with general contracting
backgrounds~\cite{tolleywesley}. And here we find that these conclusions also apply to our single-field mechanism.

Nevertheless, this instability should not be construed as a show-stopper. In the New Ekpyrotic scenario, for instance, it is circumvented by
including a pre-ekpyrotic stabilizing phase that brings the field arbitrary close to the tachyonic ridge~\cite{newekinit}.
More importantly, as we will see later on, the growth of $\zeta$ on super-Hubble scales enhances non-linearities and 
leads to interesting non-Gaussian signatures.

\section{Phenomenological Considerations} \label{4}

In this Section we describe various constraints on the scale invariant phase, ranging from requiring that the modes, generated inside the sound horizon, 
cross the Hubble horizon, to solving the standard problems of big bang cosmology, to generating acceptable levels of non-Gaussianity.
We will see that these conditions together require the background evolution to be inflationary (in the expanding case) or ekpyrotic (in the contracting case).

\subsection{Generating super-Hubble modes}
\label{suphub}

For general $\epsilon$, the amplification and freeze-out mechanism described above relies on 
a rapidly changing sound horizon. And indeed it is conceivable that the Hubble radius and
sound horizons are vastly different during this early-universe phase. But once the universe exits this phase and
becomes radiation dominated, we must ensure that we are left with scale-invariant perturbations
{\it beyond the Hubble radius} at reheating. In other words, during the generation of perturbations, modes must not only exit the sound horizon, but also the Hubble radius.

Therefore the comoving Hubble radius, $H^{-1}/a$, must be shrinking in time. From~(\ref{const}), this is given by
\be
\frac{H^{-1}}{a}\sim (-y)^{\frac{\epsilon-1}{\epsilon_s+\epsilon-1}}\,,
\ee
and thus we require that the exponent be positive. This means $\epsilon < 1$ in Case I,
corresponding to an inflationary universe, or $\epsilon > 1$ in Case II, describing a contracting universe with $w> -1/3$.
While it is of course not surprising that the expanding branch must be inflationary, we should emphasize that this is not
necessarily slow-roll inflation --- here $\epsilon$ can take on must larger values and still satisfy constraints on deviations from scale invariance. 

\subsection{Duration of Scaling Behavior} \label{dura}

Although it is traditional to require scale invariant perturbations from the largest observable scales all the way down to the reheating scale, in reality observations
only constrain near scale invariance over approximately 10 e-folds of modes, from CMB scales ($k^{-1} \sim 10^{3}$~Mpc) down to galactic scales
($k^{-1} \sim$~Mpc). Thus a conservative requirement on the duration of our scaling phase is
\begin{equation} \label{duration}
\ln\left(\frac{\left.aH/c_s\right\vert_{\rm end}}{\left.aH/c_s\right\vert_{\rm CMB}}\right) \gtrsim 10 \, .
\end{equation}
As mentioned in Sec.~\ref{ampstab}, for instance, the scaling phase could be followed by a phase of nearly constant $c_s$, which would result in a different spectral tilt on
sub-galactic scales.

\subsection{Homogeneity, Flatness and Isotropy}

We can also constrain $\epsilon$ by requiring that the background solves the flatness, horizon and isotropy problems of standard
big bang cosmology. Consider the general Friedmann equation
\be
H^2  = -\frac{K}{a^2} + \frac{8\pi G}{3}\left(\frac{C_{\rm dust}}{a^3} + \frac{C_{\rm rad}}{a^4} + \frac{C_{\rm aniso}}{a^6}\right) + \frac{8\pi G}{3}\frac{C_\phi}{a^{2\epsilon}}\,,
\label{fried}
\ee
where on the right-hand side we have allowed respectively for spatial curvature, dust, radiation and anisotropic stress contributions, as well as our general component
with equation of state parameter $\epsilon$. 

In an expanding universe, the most dangerous term is spatial curvature --- this is the flatness problem.
Hence, our extra component must redshift more slowly than $1/a^2$ to be an attractor of the cosmological
evolution. By inspection, this requires $\epsilon < 1$, or $w < -1/3$, which is again the condition for inflation.

In a contracting background, on the other hand, the worrisome component is the anisotropy, which blueshifts as $1/a^6$. Thus a contracting
universe typically becomes increasingly anisotropic, which is at the root of chaotic mixmaster behavior~\cite{BKL,chaosothers}.
In order to blueshift faster and win over anisotropy, our component must satisfy $\epsilon  > 3$, or $w > 1$~\cite{gratton,mixmaster}.
This is the ekpyrotic branch~\cite{newek,ek1,ek2,ek3,tolley,cyclic}. See~\cite{jlrev} for a review of the ekpyrotic/cyclic scenario.
Indeed, ekpyrotic contraction does equally well at solving the flatness, horizon and isotropy problems as inflationary expansion does.

Two points are worth emphasizing:

\begin{itemize}

\item The model with varying sound speed described here is the first example of a {\it single-field} ekpyrotic model resulting in an unambiguously
scale invariant spectrum for the curvature perturbation. The earliest renditions of the ekpyrotic/cyclic scenarios also invoked a single scalar field, but $\zeta$ was not
scale invariant during the contracting phase, and one had to invoke higher-dimensional effects at the singularity~\cite{tolley,other5d,lythandfriends,paolonic} or a breakdown of the
equation of motion for $\zeta$ at the bounce~\cite{robertrecent}. In New Ekpyrotic Cosmology,
this issue was resolved by using two ekpyrotic scalar fields to generate a scale-invariant entropy perturbation~\cite{newek,fabio,lehners,paolo,wands,notari}. 

\item These earlier mechanisms, either relying on matching conditions at the singularity or with two scalar fields, all require $\epsilon \gg 3$ for the spectrum to
be nearly scale invariant. Here the condition that $\epsilon> 3$ is much weaker, thereby extending the range of allowed models.

\end{itemize}

While the range $\epsilon> 3$ is certainly desirable as it allows the ekpyrotic component to dominate over all other forms of energy density,
this condition is by no means necessary from the strict point of view of generating perturbations. The actual lower bound comes
from generating super-Hubble modes which, as we have seen in Sec.~\ref{suphub}, only requires $\epsilon > 1$. 
For instance, the instability in Case II presumably requires some earlier phase to bring the field trajectory sufficiently
close to the desired classical trajectory; this pre-scaling phase could also drive the universe towards homogeneity and isotropy
using the ekpyrotic smoothing mechanism if $\epsilon > 3$.

\subsection{Constraints from Gravity Waves and non-Gaussianities}

Further constraints to our model come from the tilt  of the gravitational wave spectrum and non-Gaussianity, which will be discussed in more detail in Secs.~\ref{gw} and~\ref{running}, respectively. Interestingly, the cancelation between $\epsilon$ and $\epsilon_s$ that results in a scale invariant two-point function does not
protect the tensor spectrum nor the three-point function from running.

The inflationary branch (Case I) is characterized by a decreasing sound speed, which in turn leads to larger non-Gaussianities on small scales.
In order to maintain perturbative control over the entire range of modes, we must require $\epsilon  \lesssim 0.5$ (or $w \lesssim -2/3$). Meanwhile, the 
tensor spectrum is red and therefore peaks on large scales. To satisfy current bounds on the gravitational wave contribution to the CMB temperature anisotropy,
we must impose $\epsilon  \lesssim 0.3$.

In the ekpyrotic branch (Case II), the growing sound speed results in non-Gaussianities that grow in amplitude as we probe larger scales; but clearly the sound speed should not exceed the speed of light during the evolution.  Moreover, the growing of the modes outside the horizon generate a new plethora of terms that have to be included in the three-point function (see~\eqref{3ampII}) and that 
can very easily produce unacceptably large level of non-Gaussianity on CMB scales. However, we can still find a safe region in the $(c_{c, \rm end},\epsilon)$ parameter space. With $f_X = 1$ --- see~(\ref{fX}) for its definition---, for instance, the allowed region lies between $c_s \gtrsim 0.7$ and $2.8 \lesssim \epsilon \lesssim 3.15$, as shown in Fig.~\ref{fit1}.
 Meanwhile, the tensor spectrum has a strong blue tilt and therefore does not yield any constraint on the
equation of state parameter.

To summarize, for our two branches of solutions, corresponding respectively to expansion and contraction, we have the allowed respective range for the equation of state parameter:
\bea
\nonumber
0 \; < &\epsilon& \lesssim \; 0.3 \qquad ({\rm Case\;I})\; \\
2.8 \; \lesssim  &\epsilon & \lesssim\;  3.15 \;\;\;\;\;  ({\rm Case\;II})\,.
\label{constraintsummary}
\eea
Intriguingly, the bounds on the equation of state are comparably constraining in each case: roughly up to 20\% allowed deviations from exact de Sitter ($-1 < w \lesssim -0.8$) or kinetic domination ($0.9 < w \lesssim 1.1$), respectively. Note that in Case II the sound speed must be suitably adjusted as a function of $\epsilon$ in order to give acceptable non-Gaussianity, as shown in Fig.~\ref{fit1}. Each branch addresses the standard problems of big bang cosmology and, for suitably varying speed of sound, yields a scale invariant spectrum of density perturbations while being consistent with present limits on gravitational waves and non-Gaussianity.

\section{Scalar Spectral Index}
\label{ns}

In this Section we consider small departures from exact scale invariance, which
occur whenever $\epsilon_s$ deviates from the conditions~(\ref{caseI}) or~(\ref{caseII}),
and/or when $\epsilon,\epsilon_s$ are time-dependent. To parametrize this
time dependence, we introduce 
\be
\eta = \frac{1}{H}\frac{{\rm d}\ln\epsilon}{{\rm d}t}\;;\qquad \eta_s =  \frac{1}{H}\frac{{\rm d}\ln\epsilon_s}{{\rm d}t}\,.
\ee
For simplicity, we treat $\eta$ and $\eta_s$ as small and constant. 
Nothing else is assumed about $\epsilon$ and $\epsilon_s$ for the moment.

In this approximation, our time redefinition, ${\rm d}y = c_s {\rm d}\tau$, can be integrated explicitly:
\be \label{y}
y = \frac{c_s}{(\epsilon+\epsilon_s -1)aH}\left(1+\frac{\epsilon\eta + \epsilon_s\eta_s}{(\epsilon+\epsilon_s-1)^2}\right)\,.
\ee
And this in turn can be used to obtain an explicit expression for $q''/q$ in the mode equation~(\ref{veqn}). For instance, we have
\be
\frac{{\rm d}\ln q}{{\rm d}\ln y} = \frac{1+(\eta-\epsilon_s)/2}{\epsilon+\epsilon_s -1}\left(1+\frac{\epsilon\eta + \epsilon_s\eta_s}{(\epsilon+\epsilon_s-1)^2}\right)\,.
\ee
Proceeding along these lines, after some straightforward algebra the mode equation takes the form
\be
y^2v''_k +y^2k^2v_k - \left(\nu^2-\frac{1}{4}\right)v_k = 0\,,
\label{veqn2}
\ee
where
\bea
\nonumber
\nu^2 - \frac{1}{4}&=& \frac{1}{(\epsilon+\epsilon_s -1)^2}\left\{\left(1 +\frac{\eta - \epsilon_s}{2}\right)^2 - \frac{\epsilon_s\eta_s}{2} + \frac{(1-\epsilon_s/2)(4-3\epsilon_s-2\epsilon)(\epsilon\eta+\epsilon_s\eta_s)}{(\epsilon+\epsilon_s -1)^2}\right\} \\
&-& \frac{1 + (\eta - \epsilon_s)/2}{\epsilon+\epsilon_s -1}\,.
\label{nu}
\eea

As usual we can turn~(\ref{veqn2}) into a Bessel equation for $v_k/\sqrt{-ky}$, which, assuming adiabatic initial conditions, has a Hankel function solution, 
$H^{(1)}_\nu(-ky)$. In the long-wavelength limit, the corresponding spectral index is then
\be
n_s - 1 = -2\nu + 3\,.
\ee
Thus the spectrum is nearly scale invariant for $\nu\approx 3/2$. In particular, setting $\eta = \eta_s = 0$, one can quickly check that exact scale invariance ($\nu =3 /2$) requires
that either~(\ref{caseI}) or~(\ref{caseII}) be satisfied. We next derive explicit expressions for $n_s$ for small departures from these conditions.

\begin{itemize}

\item {\bf Case I: $\epsilon_s \approx -2\epsilon$}

It is convenient to characterize deviations from $\epsilon_s = -2\epsilon$ through a small parameter $\delta_{\rm I}$:
\be \label{deltaI}
\epsilon_s = -2\epsilon + (1+\epsilon)\delta_{\rm I}\,.
\ee
By expanding $\nu$ to linear order in $\delta_{\rm I}$, $\eta$, and $\eta_s$, the scalar spectral index is thus given by
\be
\left(n_s - 1\right)_{\rm I} = -\delta_{\rm I} - \frac{1}{(1+\epsilon)^2}\left[\left(1+\frac{11}{3}\epsilon\right) \eta - \frac{14}{3} \epsilon\eta_s\right]\,.
\ee
As a check, note that in the limit of quasi-de Sitter background and nearly constant $c_s$, for which $\epsilon,\epsilon_s\ll 1$, we have
$\delta_{\rm I} \approx \epsilon_s + 2\epsilon$, and thus 
\be
\left(n_s - 1\right)_{\rm slow-roll} \approx -2\epsilon - \epsilon_s - \eta\,.
\ee
This agrees with earlier treatments~\cite{garrigamukhanov}. 

More generally, however, larger values of $\epsilon$ and $\epsilon_s$ are possible
without spoiling approximate scale invariance, provided $\delta_{\rm I}$ remains small. Thus the $\delta_{\rm I}$ parametrization offers a more general
characterization of deviations from $n_s = 1$, which, in the special case $\epsilon,\epsilon_s\ll 1$, reproduces standard slow-roll results.
 
Instead of $\eta_s$, we can alternatively introduce a parameter $\delta'_{\rm I}$ obtained by taking the derivative of~\eqref{deltaI}:
\begin{equation}
\delta'_{\rm I}\equiv \frac{{\rm d} \delta_{\rm I}}{{\rm d} \log a}\,.
\end{equation}
Expressed in terms of this parameter, we find
\be
\left(n_s - 1\right)_{\rm I} = -\delta_{\rm I} - \frac{1-\epsilon}{(1+\epsilon)^2}\eta -\frac{7}{3}\frac{\delta'_{\rm I}}{1+\epsilon}\,.
\ee

\item {\bf Case II: $\epsilon_s \approx \frac{2}{5}(3-2\epsilon)$}

Similarly, in the contracting branch we can introduce $\delta_{\rm II}$ as
\be
\epsilon_s = \frac{2}{5}(3-2\epsilon) - \frac{1}{5}(1+\epsilon)\delta_{\rm II}\,.
\ee
Substituting into~(\ref{nu}) and, as before, keeping terms linear in $\delta_{\rm II}$, $\eta$, and $\eta_s$, we obtain
\be
\left(n_s - 1\right)_{\rm II} = -5\;\delta_{\rm II} -\frac{5}{(1+\epsilon)^2}\left[ \left(1+\frac{43}{3}\epsilon\right)\eta + 70\left(1-\frac{2}{3}\epsilon\right )\eta_s\right]\,.
\ee
Or, equivalently, in terms of $\delta'_{\rm II}\equiv {\rm d} \delta_{\rm II}/{\rm d} \log a$,
\be
\left(n_s - 1\right)_{\rm II} = -5\;\delta_{\rm II} -\frac{5}{(1+\epsilon)^2}\left[ \left(1-\frac{97}{3}\epsilon\right)\eta - \frac{35}{3}(1+\epsilon)\delta'_{\rm II}\right]\,.
\ee
\end{itemize}

In slow-roll inflation $\epsilon$ and $\eta$ are usually considered as independent and comparably small parameters. Similarly, here we treat $\delta$, $\eta$ and $\eta_s$ (or $\delta$, $\eta$ and $\delta'$) on the same footing because that all appear at linear order in the expression of the observationally small quantity $n_s-1$. Of course, more fine tuned cancellations and richer hierarchies between the parameters can also be considered.

\section{Gravitational Wave Spectrum}
\label{gw}

Unlike scalar perturbations, tensor modes are oblivious to a time-dependent sound speed and instead feel only the cosmological background. The mode functions $h_k(\tau)$
therefore satisfy
\be
\frac{{\rm d}^2h_k}{{\rm d}\tau^2} + \left(k^2 -  \frac{2-\epsilon}{(1-\epsilon)^2\tau^2}\right)h_k = 0\,.
\ee
As usual, this can be cast as a Bessel equation through the field redefinition: $u_k = ah_k/\sqrt{-k\tau}$. Assuming adiabatic initial conditions, the solutions are given by
\be
h_k(\tau) = \frac{\sqrt{-\tau\pi}}{2a}H_{\frac{1}{2}\left\vert\frac{3-\epsilon}{1-\epsilon}\right\vert}^{(1)}(-k\tau)\,.
\ee

\begin{itemize}

\item {\bf Case I: $\epsilon_s \approx -2\epsilon$}

In the inflationary branch, the equation of state parameter falls in the range $0<\epsilon < 1$. The tensor power spectrum, $P_h = k^3|h_k|^2/\pi^2$, therefore takes the form
\be
P_h^{({\rm I})} \sim \frac{H_{\rm end}^2}{M_{\rm Pl}^2}\left(\frac{k}{H_{\rm end}}\right)^{-\frac{2\epsilon}{1-\epsilon}}\,,
\label{PhI}
\ee
corresponding to a red tensor tilt:
\be
n_{\rm T}^{({\rm I})} = \frac{{\rm d}\ln P_h(k)}{{\rm d}\ln k} = -\frac{2\epsilon}{1-\epsilon}\,.
\ee

The tensor-to-scalar ratio follows by comparing with~(\ref{PI}). Noting that the sound speed evaluated at horizon crossing satisfies $\bar{c}_s \sim k^{-2\epsilon/(1+\epsilon)}$, we obtain
\be
r^{({\rm I})} \sim  \epsilon\bar{c}_s\left(\frac{K_{\rm end}}{K}\right)^{\frac{4\epsilon^2}{1-\epsilon^2}}\,.
\ee
And since this amplitude peaks on the largest scales, we have to ensure that this satisfies current bounds on the gravitational wave contribution to the CMB: $r_{\rm CMB} \lesssim 0.3$. We will see in Sec.~\ref{ngsign} that the non-Gaussian amplitude is proportional to $\bar{c}_s^{-2}$, and thus the current constraint from non-Gaussianity gives $\bar{c}_s\vert_{\rm CMB}\;\gsim \; 0.1$. Moreover, as discussed in Sec.~\ref{dura}, we have $K_{\rm end}/K\vert_{\rm CMB}\;\gsim\; 10^3$. Assuming the minimum required number of e-folds, we obtain the conservative bound
\be
\epsilon \lesssim 0.3\,.
\ee
This confirms the bound quoted in~(\ref{constraintsummary}).

\item {\bf Case II: $\epsilon_s \approx \frac{2}{5}(3-2\epsilon)$}

In the ekpyrotic branch, the gravitational wave power spectrum is
\be
P_h^{({\rm II})} \sim  \frac{H_{\rm end}^2}{M_{\rm Pl}^2}\left(\frac{k}{H_{\rm end}}\right)^{3 - |\frac{\epsilon-3}{\epsilon-1}|}\,,
\ee
For $\epsilon> 3$ the tensor spectrum has a strong blue tilt:
\be
 n_{\rm T}^{({\rm II})} = \frac{2\epsilon}{\epsilon-1} > 2\,.
\ee
In particular, the gravitational wave amplitude is negligible on CMB scales in this case, as in the pre-big bang~\cite{pbb,pbbgw} and ekpyrotic scenarios~\cite{ek1,gwaves}. For $\epsilon<3$ we have 
\be
 n_{\rm T}^{({\rm II})} = \frac{4\epsilon-6}{\epsilon-1}.
\ee
The gravitational wave spectrum is flat for $\epsilon = 3/2$, {\it i.e.}, in the case of dust-like contraction~\cite{dust}.

\end{itemize}

\section{Calculating Non-Gaussianities}
\label{7}

We now calculate the three-point function for the following general single scalar field model~\cite{chen,seery}:
\begin{equation} \label{general}
S=\int {\rm d}^4x \sqrt{-g} \left[\frac{R}{2} +
P(X,\phi)\right]~,
\end{equation}
where  the pressure $P$ is a general function of the scalar field $\phi$ and the kinetic term $X=-\frac{1}{2}g^{\mu\nu}\partial_{\mu}\phi \partial_{\nu}\phi$. To simplify the expressions below, 
we set $M_{\rm Pl} = 1$ throughout the calculation. The energy density reads 
\begin{equation}
\rho = 2 X P_{,X} - P\,,
\end{equation}
while the speed of sound is given by
\begin{equation} 
c_s^2 = \frac{P_{,X}}{\rho_{,X}}= \frac{P_{,X}}{P_{,X}+2X P_{,XX}}\,.
\end{equation}
Besides the parameters $\epsilon$, $\es$, $\eta$, $\eta_s$ introduced earlier, the calculation of the three-point function also necessitates defining two further parameters derived
from $P(X,\phi)$~\cite{seery,chen}
\begin{eqnarray}
\Sigma&=&X P_{,X}+2X^2P_{,XX}  = \frac{H^2\epsilon}{c_s^2} ~,\\
\lambda&=& X^2P_{,XX}+\frac{2}{3}X^3P_{,XXX} ~.
\label{lambda}
\end{eqnarray}

Consider for instance the DBI effective action~\cite{eva} describing the motion of a 3-brane in a throat of some
higher-dimensional compact space:
\be
P_{\rm DBI}(X,\phi) = -f^{-1}(\phi)\sqrt{1-2f(\phi)X}+f^{-1}(\phi)-V(\phi)\,,
\label{DBI}
\ee
where $f(\phi)$ encodes the warp factor of the throat (see e.g. \cite{chris} for non-standard throat geometries). In this case we have
\bea
c_s^2 = 1-2f(\phi)X\,; \qquad \Sigma = \frac{X}{(1-2f(\phi)X)^{3/2}}\,;\qquad \lambda = \frac{X^2f(\phi)}{(1-2f(\phi)X)^{5/2}}\,.
\label{DBIquantities}
\eea
Although our calculation of non-Gaussiniaties below applies to general $P(X,\phi)$, we view the DBI action as a fiducial model for our analysis. 
In particular, in Appendix \ref{appendixA} we follow~\cite{kinney} and provide explicit examples of DBI Lagrangians that yield scaling solutions of the form~(\ref{const}).

In the present case, the cubic effective action has already been derived in~\cite{chen,seery}.
The result is valid outside of the slow-roll approximation and for any time-dependent sound speed:
\begin{eqnarray} \label{action3}
S_{(3)}&=& M_{\rm Pl}^2 \int {\rm d}t {\rm d}^3x \left\{
-a^3 \left[\Sigma\left(1-\frac{1}{c_s^2}\right)+2\lambda\right] \frac{\dot{\zeta}^3}{H^3}
+\frac{a^3\epsilon}{c_s^4}(\epsilon-3+3c_s^2)\zeta\dot{\zeta}^2 \right.
\nonumber \\ &+&
\frac{a\epsilon}{c_s^2}(\epsilon-2\es+1-c_s^2)\zeta(\partial\zeta)^2-
2a \frac{\epsilon}{c_s^2}\dot{\zeta}(\partial
\zeta)(\partial \chi) \nonumber \\ &+& \left.
\frac{a^3\epsilon}{2c_s^2}\frac{d}{dt}\left(\frac{\eta}{c_s^2}\right)\zeta^2\dot{\zeta}
+\frac{\epsilon}{2a}(\partial\zeta)(\partial
\chi) \partial^2 \chi +\frac{\epsilon}{4a}(\partial^2\zeta)(\partial
\chi)^2+ 2 f(\zeta)\left.\frac{\delta L_{(2)}}{\delta \zeta}\right\vert_1 \right\} ~,
\end{eqnarray}
where dots denote derivatives with respect to proper time $t$, $\partial$ is a spatial derivative,
and $\chi$ is defined as
\begin{equation}
\partial^2 \chi = \frac{a^2 \epsilon}{c_s^2}\dot{\zeta}\,.
\end{equation}
Meanwhile, in the last term $\frac{\delta L_{(2)}}{\delta\zeta}|_1$ denotes the variation of the
quadratic action with respect to the perturbation $\zeta$:
\begin{eqnarray}
\left.\frac{\delta
L_{(2)}}{\delta\zeta}\right\vert_1 &=& a
\left( \frac{d\partial^2\chi}{dt}+H\partial^2\chi
-\epsilon\partial^2\zeta \right) ~,
\end{eqnarray}
\begin{eqnarray} \label{redefinition}
f(\zeta)&=&\frac{\eta}{4c_s^2}\zeta^2+\frac{1}{c_s^2H}\zeta\dot{\zeta}+
\frac{1}{4a^2H^2}[-(\partial\zeta)(\partial\zeta)+\partial^{-2}(\partial_i\partial_j(\partial_i\zeta\partial_j\zeta))] \nonumber \\
&+&
\frac{1}{2a^2H}[(\partial\zeta)(\partial\chi)-\partial^{-2}(\partial_i\partial_j(\partial_i\zeta\partial_j\chi))] ~,
\end{eqnarray}
where $\partial^{-2}$ is the inverse Laplacian. In order to calculate the three-point function in the interaction picture we can derive the Hamiltonian $H_{\rm int}$ in the usual way. Except for the last term in \eqref{action3} we have 
$H_{\rm int} = - L_{(3)}$, up to interactions that are higher order in the number of fields. The last term requires more attention because it contains a second time derivative in the field which makes problematic the definition of the conjugate momentum. However, since $\frac{\delta L}{\delta\zeta}|_1$ is proportional to
the linearized equations of motion, it can be absorbed by a field redefinition
\begin{eqnarray}
\zeta \rightarrow \zeta_n+f(\zeta_n) ~.
\end{eqnarray}

Only the first two terms in~(\ref{redefinition}) contribute to the three-point correlation function at long wavelengths since
all other terms involve gradients that go to zero outside the horizon. In Case I, the second term is also negligible since
$\zeta\rightarrow {\rm const.}$ on super Hubble scales. Thus, after the field redefinition $\zeta \rightarrow \zeta_n+\frac{\eta}{4c_s^2}\zeta_n^2$,
the three-point function can be expressed in this case as
\begin{eqnarray}
\langle\zeta(\textbf{x}_1)\zeta(\textbf{x}_2)\zeta(\textbf{x}_3)\rangle_{\rm (\bf I)}
&=&\langle\zeta_n(\textbf{x}_1)\zeta_n(\textbf{x}_2)\zeta_n(\textbf{x}_3)\rangle
\nonumber \\
&+&\frac{\eta}{2c_s^2}
\left(\langle\zeta_n(\textbf{x}_1)\zeta_n(\textbf{x}_2)\rangle
\langle\zeta_n(\textbf{x}_1)\zeta_n(\textbf{x}_3)\rangle+\textrm{cyclic}\right) + {\cal O} (\zeta_n^6)\,. 
\end{eqnarray}
In Case II, on the other hand, the second term can be simplified by noting that $\zeta\sim 1/y^3$ on super-Hubble scales. To leading order in $\eta$,$\eta_s$, it
follows from~(\ref{y}) and the condition for scale invariance~(\ref{caseII}) that $\dot{\zeta}_n = -(3/5)H(1+\epsilon)\zeta_n$.
Thus the desired field redefinition is $\zeta \rightarrow \zeta_n+\frac{\eta}{4c_s^2}\zeta_n^2 - \frac{3(1+\epsilon)}{5c_s^2}\zeta_n^2$, in terms of which the
three-point amplitude takes the form
\begin{eqnarray}
\langle\zeta(\textbf{x}_1)\zeta(\textbf{x}_2)\zeta(\textbf{x}_3)\rangle_{\rm (\bf II)}
&=&\langle\zeta_n(\textbf{x}_1)\zeta_n(\textbf{x}_2)\zeta_n(\textbf{x}_3)\rangle
\nonumber \\
\nonumber
&+&\left(\frac{\eta}{2c_s^2} - \frac{6(1+\epsilon)}{5c_s^2} \right)
\left(\langle\zeta_n(\textbf{x}_1)\zeta_n(\textbf{x}_2)\rangle
\langle\zeta_n(\textbf{x}_1)\zeta_n(\textbf{x}_3)\rangle+\textrm{cyclic}\right) \\
&+& {\cal O} (\zeta_n^6)\,. 
\label{fredef}
\end{eqnarray}

\subsection{The three-point function: some useful formulae}

We now proceed to calculate the three-point function using standard methods \cite{malda}. At first order in perturbation theory and in the interaction picture we have 
\begin{equation} \label{interaction}
\langle
\zeta(t,\textbf{k}_1)\zeta(t,\textbf{k}_2)\zeta(t,\textbf{k}_3)\rangle=
-i\int_{t_0}^{t}{\rm d}t^{\prime}\langle[
\zeta(t,\textbf{k}_1)\zeta(t,\textbf{k}_2)\zeta(t,\textbf{k}_3),H_{\rm int}(t^{\prime})]\rangle ~,
\end{equation}
where $t_0$ is some sufficiently early time. It proves convenient to work with the sound horizon time variable $y$ defined in \eqref{y} (${\rm d}y = c_s {\rm d}t/a$) and expand the field $\zeta$ in creators and annihilators in the usual way:
\begin{equation} \label{modes} 
\zeta(y, \kk) = u_k(y)a(\kk) + u_k^*(y) a^\dagger(-\kk) .
\end{equation}
Note that with our Fourier conventions commutation relations read
$[a(\kk), a^\dagger(\kk')] = (2 \pi)^3 \delta^3(\kk - \kk')$.
The mode functions $u_k$ are easily computed in terms of the canonically normalized mode functions $v$~\eqref{vsol}. Up to an irrelevant phase, we have
\begin{equation} \label{u}
u_k(y) = \frac{1}{a}\left(\frac{c_s}{2 \epsilon}\right)^{1/2} v_k(y) = 
\frac{H \comb}{2 \sqrt{c_s k^3 \epsilon }}(1 + i k y) e^{-i k y}.
\end{equation}

The contribution to the three-point function of each term in~\eqref{action3} can be worked out once we substitute~\eqref{modes} and~\eqref{u} in~\eqref{interaction}. After switching to the integration variable $y$ and making some simplifications, we encounter integrals of the form 
\begin{equation} \label{integral}
{\cal C} = \int_{-\infty + i\varepsilon}^{y_{\rm end}} {\rm d} y \left(\frac{y}{y_{\rm end}}\right)^\gamma (- i y)^n e^{i K y}\, .
\end{equation}
Only the imaginary part of this integral is relevant for our calculation.
In the above expression, $n$ is an integer, $y_{\rm end}<0$ is the value of $y$ at the end of the inflationary or ekpyrotic phase, and $K \equiv k_1+k_2+k_3$. Hence, for all modes of interest, $K |y_{\rm end}|$ is a small quantity.  
As discussed in~\cite{malda}, the above choice of integration contour picks up the appropriate interacting ``in-in" vacuum at $|y|\rightarrow \infty$ and takes care of the oscillating behavior of the exponential.

For simplicity, we consider only strictly scale invariant solutions and therefore neglect small corrections of order $\delta$, $\eta$ and $\eta_s$ in our calculations. The exponent $\gamma$ is thus a constant, parameterizing the behavior of either $H/c_s^{1/2}$ or 
$H/c_s^{5/2}$, depending on the term being considered. 
More specifically, in Case I (expanding branch) we have
\begin{equation} \label{0alpha}
\frac{H}{c_s^{1/2}} = \frac{H_{\rm end}}{{\bar c}_{s\, \rm end}^{1/2}}\quad(\gamma = 0)\,, \quad \qquad \frac{H}{c_s^{5/2}} = \frac{H_{\rm end}}{c_{s\, \rm end}^{5/2}} \left(\frac{y}{{y_{\rm end}}}\right)^\alpha \quad (\gamma=\alpha)\,,
\end{equation}
where 
\begin{equation}
\alpha \equiv -\frac{4 \epsilon}{1 + \epsilon}\,.
\label{alpdef}
\end{equation}
In Case II, instead,
\begin{equation} \label{-3beta}
\frac{H}{c_s^{1/2}} = \frac{H_{\rm end}}{{\bar c}_{s\, \rm end}^{1/2}} \left(\frac{y}{{y_{\rm end}}}\right)^{-3} \quad(\gamma = -3)\,, \quad \qquad \frac{H}{c_s^{5/2}} = \frac{H_{\rm end}}{c_{s\, \rm end}^{5/2}} \left(\frac{y}{{y_{\rm end}}}\right)^\beta \quad (\gamma=\beta)\,,
\end{equation}
where
\begin{equation}
\beta \equiv 5 \frac{\epsilon -3}{\epsilon +1}\,.
\label{betdef}
\end{equation}

Let us describe the integrals in more detail for Case I.
For $\gamma + n >-2$ the imaginary part of \eqref{integral} is convergent as $y_{\rm end}\rightarrow 0$. In this case we can approximately extend the upper limit of integration to $0$, which amounts to neglecting terms of higher order in $k |y_{\rm end}|$. We thus obtain
\begin{equation} \label{Cconvergent}
{\rm Im}\, {\cal C} = - (K |y_{\rm end}|)^{-\gamma} \cos\frac{\gamma \pi}{2} \Gamma(1+ \gamma + n) K^{-n-1}\, .
\end{equation}
When $\gamma+n<-2$, however, the integral \eqref{integral} is divergent for $y_{\rm end} \rightarrow 0$. To leading order in  $k |y_{\rm end}|$ we find
\begin{equation} \label{Cdivergent}
{\rm Im}\, {\cal C} = {\rm Im} \frac{(-i)^n}{\gamma+n+1}(y_{\rm end}^{n+1} + i K y_{\rm end}^{n+2})\,.
\end{equation}
For example, 
\begin{equation}
{\rm Im}\, {\cal C} = \frac{(K |y_{\rm end}|)^2}{K(\gamma+2)} \quad (n=0)\,,\qquad
{\rm Im}\, {\cal C} = - \frac{(K |y_{\rm end}|)^2}{K^2(\gamma+2)} \quad (n=1)\,,
\end{equation}
and so forth. For the parameter range relevant to Case II the integrals are calculated in  Appendix \ref{appendixB}.

\subsection{Example 1: the $\zeta \dot\zeta^2$ contribution}

Before writing our full set of results, it is worth outlining as an example the calculation of the 3-point contribution from the $\zeta \dot\zeta^2$ term in~\eqref{action3}. By applying the commutation relations this reads:
\begin{multline}
\langle \zeta(\textbf{k}_1)\zeta(\textbf{k}_2)\zeta(\textbf{k}_3)\rangle_{\zeta \dot\zeta^2}\, =\,
i (2 \pi)^3 \delta^3(\kk_1+\kk_2+\kk_3) u_{k_1}(y_{\rm end})u_{k_2}(y_{\rm end})u_{k_3}(y_{\rm end}) \\ \times 
\int_{-\infty+i\varepsilon}^{y_{\rm end}} {\rm d} y \frac{c_s}{a} \frac{a^3 \epsilon}{c_s^4} (\epsilon - 3 + 3 c_s^2)
u_{k_1}^*(y) \frac{d u_{k_2}^*(y)}{d y}\frac{d u_{k_3}^*(y)}{d y} + {\rm perm.} + {\rm c.c.}
\end{multline}
Substituting~\eqref{u}, we have
\begin{multline}
\langle \zeta(\textbf{k}_1)\zeta(\textbf{k}_2)\zeta(\textbf{k}_3)\rangle_{\zeta \dot\zeta^2}\, =\, i (2 \pi)^3 \delta^3(\kk_1+\kk_2+\kk_3) 
\frac{H_{\rm end}^3 \comb^3}{4^3\, \epsilon^{\, 3/2}\, c_{s\, \rm end}^{\, 3/2}}\frac{1}{\Pi_j k_j^3}  \\
\times \int_{-\infty+i\varepsilon}^{{y_{\rm end}}} {\rm d}y \frac{a^2 \epsilon}{c_s^3} (\epsilon - 3 + 3 c_s^2)
\frac{{H}^3 \comb^3}{\epsilon^{\, 3/2}\, {c}_s^{\, 3/2}} (1 - i k_1 y) k_2^2 k_3^2 y^2 e^{i K y}  + {\rm perm.} + {\rm c.c.}
\end{multline}
Since we are considering a strictly scale-invariant solution, $\epsilon$ and $\epsilon_s$ are constant and can be taken outside of the integral. Now, we can use~\eqref{y} to substitute for $a^2 y^2$  and find
\begin{multline}
\langle \zeta(\textbf{k}_1)\zeta(\textbf{k}_2)\zeta(\textbf{k}_3)\rangle_{\zeta \dot\zeta^2}\, =\, i (2 \pi)^3 \delta^3(\kk_1+\kk_2+\kk_3) 
\frac{ H_{\rm end}^3 \comb^4}{4^3\, \epsilon^{2}\, { c}_{s\, \rm end}^{\, 3/2}}\frac{1}{\Pi_j k_j^3}  \\ \times \int_{-\infty+i\varepsilon}^{{y_{\rm end}}} {\rm d}y \frac{H}{c_s^{5/2}} (\epsilon - 3 + 3 c_s^2)
(1 - i k_1 y) k_2^2 k_3^2 e^{i K y}  + {\rm perm.} + {\rm c.c.}
\end{multline}

In the expanding case (Case I), $\epsilon_s=-2\epsilon$, and the background-dependent functions in the integrand display the two types of time dependence listed in~\eqref{0alpha}. Since both of these give rise to convergent integrals, we can apply~\eqref{Cconvergent}. Moreover, we combine the prefactor $(K |y_{\rm end}|)^{-\gamma}$ with the appropriate background functions, $H_{\rm end}/c^{1/2}_{s\, \rm end}$ or $H_{\rm end}/c^{5/2}_{s\, \rm end}$, to express the corresponding quantities at horizon crossing. In other words, $H_{\rm end}c^{-1/2}_{s\, \rm end} = {\bar H}{\bar c}^{-1/2}_{s}$, $H_{\rm end}c^{-5/2}_{s\, \rm end} (K |y_{\rm end}|)^{-\alpha} = {\bar H}{\bar c}^{-5/2}_{s}$
where bar quantities are evaluated at horizon crossing. We finally obtain (Case I)
\begin{multline}
\langle \zeta(\textbf{k}_1)\zeta(\textbf{k}_2)\zeta(\textbf{k}_3)\rangle_{\zeta \dot\zeta^2\ (\bf I)}\, =\, (2 \pi)^3 \delta^3(\kk_1+\kk_2+\kk_3) 
\frac{{\bar H}^4 \combtwo^4}{16 \epsilon^{2}\, {\bar c}_s^{4}}\frac{1}{\Pi_j k_j^3} 
\frac{k_2^2 k_3^2}{K} \\
\times \left\{(\epsilon-3) \cos\frac{\alpha \pi}{2} \Gamma(1+\alpha)\left[1+ (1+ \alpha) \frac{k_1}{K}\right] + 3 {\bar c}_s^2 \left[1+ \frac{k_1}{K}\right] \right\}+{\rm sym}.
\end{multline}
Note that the combination $\cos\frac{\alpha \pi}{2} \Gamma(1+\alpha)$, which appears ubiquitously in Case I (see below), is regular in the limit $\alpha\rightarrow -1$: the constraint $\alpha \gtrsim -1$ ($\epsilon \lesssim 0.3$) is uniquely dictated by phenomenology. 

In the contracting case (Case II), $\epsilon_s=2(3 - 2\epsilon)/5$, and the background-dependent functions in the integrand follow the power-law behaviors given in~\eqref{-3beta}. Moreover, since the amplitudes of the perturbations keep growing outside the horizon,
 it is convenient to refer to the background quantities at the end of the ekpyrosis phase as those directly relate to the power spectrum. 
 This growth of the amplitude complicates the calculation somewhat compared to the expanding case. Specifically,  the time-derivative of the mode function \eqref{u} in Case II reads 
\be
\frac{du^\dagger}{dy} = \frac{H(1-\epsilon-\epsilon_s)}{2\sqrt{c_s k^3\epsilon}}\left[k^2y -\frac{3}{y}(1-iky)\right]e^{iKy}\, .
\ee
Note that only the first term in square brackets appears in Case I. The presence of the extra terms results in longer expressions for the amplitudes in Case II. The final answer is given in Sec. \ref{allterms}.

\subsection{Example 2: the $\dot \zeta^3$ contribution}

As another example of our method, consider the $\dot{\zeta}^3$ term, for which the three-point function reads
\begin{multline}
\langle \zeta(\textbf{k}_1)\zeta(\textbf{k}_2)\zeta(\textbf{k}_3)\rangle_{\dot\zeta^3}\, =\,
- i (2 \pi)^3 \delta^3(\kk_1+\kk_2+\kk_3) u_{k_1}(y_{\rm end})u_{k_2}(y_{\rm end})u_{k_3}(y_{\rm end}) \\ \times 
\int_{-\infty+i\varepsilon}^{y_{\rm end}} {\rm d} y \frac{c_s^2}{a^2} \frac{a^3}{H^3} [\Sigma(1-c_s^{-2}) + 2 \lambda]
\frac{d u_{k_1}^*(y)}{dy} \frac{d u_{k_2}^*(y)}{d y}\frac{d u_{k_3}^*(y)}{d y} + {\rm perm.} + {\rm c.c.}
\end{multline}
Note that the parameter $\lambda$ can be written as~\cite{seery}
\begin{equation}
\lambda = \frac{\Sigma}{6}\left(\frac{2f_X}{c_s^2} + \frac{1}{c_s^2} -1\right)\, ,
\end{equation}
where we have introduced
\begin{equation}
f_X = \frac{\epsilon\epsilon_s}{3\epsilon_X}\,, 
\label{fX}
\end{equation}
with $\epsilon_X$ the ``kinetic part'' of $\epsilon$:
\begin{equation}
\epsilon = \epsilon_\phi + \epsilon_X\,; \qquad \epsilon_\phi = -\frac{\dot \phi}{H^2}\frac{\partial H}{\partial \phi}\,; \qquad \epsilon_X = -\frac{\dot X}{H^2}\frac{\partial H}{\partial X}\,.
\end{equation}
For simplicity we will henceforth assume that  $\epsilon_X$ is constant. 
Note that for the DBI action~(\ref{DBI})
\be
f_X^{\rm DBI}  = 1-c_s^2\,,
\ee
which therefore lies between 0 and 1. It follows that the $\dot{\zeta}^3$ contribution vanishes identically for the DBI case~\cite{chen}.

Explicitly, in the expanding case, we obtain
\begin{multline}
\langle \zeta(\textbf{k}_1)\zeta(\textbf{k}_2)\zeta(\textbf{k}_3)\rangle_{\dot\zeta^3\ (\bf I)}\, =\, -(2 \pi)^3 \delta^3(\kk_1+\kk_2+\kk_3) 
\frac{{\bar H}^4 \combtwo^5}{8 \epsilon^{2}\, {\bar c}_s^{4}}\frac{1}{\Pi_j k_j^3} 
\frac{k_1^2 k_2^2 k_3^2}{K^3} \times \\
\left[2 {\bar c}_s^{2} +
\left(f_X-1 \right) \cos\frac{\alpha \pi}{2} \Gamma(3 + \alpha)\right]\, .
\end{multline}
Again, for Case II, we refer to Sec. \ref{allterms}.

\subsection{Combining all terms}
\label{allterms}

The remaining contributions are calculated in a similar way to the above examples. After combining all the terms, it is convenient to express
the full three-point function by factoring out appropriate powers of the power spectrum and defining an amplitude ${\cal A}$ as
\begin{equation}
\langle \zeta(\textbf{k}_1)\zeta(\textbf{k}_2)\zeta(\textbf{k}_3)\rangle = (2\pi)^7 
\delta^3(\kk_1+\kk_2+\kk_3) P_\zeta^{\;2} \frac{1}{\Pi_j k_j^3}{\cal A}\,,
\end{equation}
where $P_\zeta$ is given in~(\ref{PI}) and~(\ref{PII}) for Cases I and II, respectively. Collecting all terms, we obtain:\\

\noindent {\bf $\bullet$ Case I:}

\begin{align} 
\nonumber {\cal A}^{(\bf I)}_{\dot\zeta^3}\, =&\, - 
\frac{1+\epsilon}{2 {\bar c}_s^{2}} 
\left[2 {\bar c}_s^{2} +
\left(f_X-1\right) \cos\frac{\alpha \pi}{2} \Gamma(3 + \alpha)\right]\frac{k_1^2 k_2^2 k_3^2}{K^3}\;; \\[2.5mm] \nonumber
{\cal A}^{(\bf I)}_{\zeta \dot\zeta^2}\, =&\, 
\frac{1}{4 {\bar c}_s^{2}} \left\{ \left[6 {\bar c}_s^2 +(\epsilon - 3) \cos\frac{\alpha \pi}{2} (2+\alpha)\Gamma(1+\alpha)\right] \frac{1}{K}\sum_{i<j}k_i^2k_j^2 +\right. \\  &\left.
- \left[3 {\bar c}_s^2+ (\epsilon - 3) \cos\frac{\alpha \pi}{2} \Gamma(2+\alpha)\right] \frac{1}{K^2} \sum_{i\neq j} k_i^2 k_j^3\right\} \nonumber\;; \\[2.5mm] \nonumber
{\cal A}^{(\bf I)}_{\zeta(\partial \zeta)^2} =&\frac{(1+5\epsilon)}{8 {\bar c}_s^2} \cos\frac{\alpha\pi}{2} \Gamma(1+\alpha)\left[\frac{1+\alpha}{1-\alpha} \sum_j k_j^3 +\frac{2-\alpha^2}{1-\alpha}\frac{2}{K}\sum_{i<j}k_i^2k_j^2 \right.\\ 
 &  -2 (1+ \alpha) \frac{1}{K^2} \sum_{i\neq j} k_i^2 k_j^3 - \frac{\alpha}{K(1-\alpha)} \sum_i k_i^4 \nonumber  \left. +\alpha \frac{1+\alpha}{1-\alpha} k_1 k_2 k_3 \right]  \\
 & - \frac{1}{8}\left[\sum_j k_j^3 + \frac{4}{K}\sum_{i<j}k_i^2k_j^2 -\frac{2}{K^2} \sum_{i\neq j} k_i^2 k_j^3 \right]\;; \nonumber \\[2.5mm]
{\cal A}^{(\bf I)}_{\dot \zeta \partial \zeta \partial \chi} =& - \nonumber
\frac{\epsilon}{4 {\bar c}_s^{2}}\cos\frac{\alpha \pi}{2}\Gamma(1+\alpha) \left[\sum_j k_j^3  + \frac{\alpha -1}{2}\sum_{i\neq j}k_i k_j^2 - 2 \frac{1+\alpha}{K^2}\sum_{i\neq j}k_i^2 k_j^3 - 2 \alpha k_1 k_2 k_3\right]\;; \\[2.5mm]
{\cal A}^{(\bf I)}_{\epsilon^2} = & \frac{\epsilon^2}{16 {\bar c}_s^2} \cos\frac{\alpha \pi}{2} \Gamma(1+\alpha)(2+\alpha/2)\left[\sum_j k_j^3 - \sum_{i\neq j} k_i k_j^2 + 2 k_1 k_2 k_3\right]\,,
\label{3ampI}
\end{align}
where ${\cal A}_{\epsilon ^2}$ accounts for the $\partial \zeta \partial \chi \partial^2 \chi$ and  $(\partial^2 \zeta) (\partial \chi)^2$ terms. \\

\noindent {\bf $\bullet$ Case II:}

\bea
\nonumber
{\cal A}^{(\bf II)}_{\dot\zeta^3} &=&  \frac{(1+\epsilon)(f_X-1)}{10 c_{s\;{\rm end}}^{2}}\left\{\frac{\cos\frac{\pi\beta}{2}}{(K|y_{\rm end}|)^\beta}\left[\Gamma(\beta+3)\frac{k_1^2k_2^2k_3^2}{K^3} \nonumber
+ 3\Gamma(\beta+2)\frac{k_1k_2k_3}{K^2}\sum_{i<j}k_ik_j \right.\right. \\
\nonumber
&+ & \left.\left. 3\Gamma(\beta+1) \left(\frac{1}{K} \sum_{i<j} k_i^2 k_j^2 + 3 k_1 k_2 k_3\right)+  9\frac{\Gamma(\beta+1)}{\beta-1}K \sum_{i<j} k_i k_j \right] \right. \\ \nonumber
&+ & \left. \frac{9K^3}{(\beta-2)(\beta-3)}\left[\frac{\cos\frac{\pi\beta}{2}}{(K|y_{\rm end}|)^\beta}\Gamma(\beta)\frac{\beta^2-2\beta+3}{\beta-1} + \frac{3}{\beta}\right]  \right\}
 \,.\nonumber
\label{zetadot3b}
\eea

\bea
\nonumber
{\cal A}^{(\bf II)}_{\zeta\dot\zeta^2} &=& \frac{(\epsilon-3)}{4 c_{s\;{\rm end}}^{2}}\left\{\frac{\cos\frac{\pi\beta}{2}}{(K|y_{\rm end}|)^\beta}
\left[\Gamma(\beta+1)\left(\frac{2+\beta}{K}\sum_{i<j}k_i^2k_j^2 -\frac{1+\beta}{K^2}\sum_{i\neq j}k_i^2k_j^3+6k_1k_2k_3\right) \right.\right. \\
\nonumber
&+ & \left.\left. 3\Gamma(\beta) \left(2\sum_{i\neq j}k_ik_j^2+9k_1k_2k_3 + \frac{9K^3}{(\beta-2)(\beta-3)}\right) \right. \right. \\
\nonumber
&+& \left.\left. 3\Gamma(\beta-1)K\left(2K^2+5\sum_{i<j} k_ik_j\right) \right] \right. \nonumber \\
\nonumber
&+ & \left. \frac{3}{\beta}\left(2K\sum_{i<j}k_ik_j +\sum_{i\neq j}k_ik_j^2 - K^3\frac{\beta^2-5\beta-3}{(\beta-2)(\beta-3)}\right) \right\} + \frac{3}{4} \sum_j k_j^3 \nonumber\,. \\
\nonumber
{\cal A}^{(\bf II)}_{\zeta(\vec{\nabla}\zeta)^2} &=& \frac{ 13 \epsilon-7}{40 c_{s\;{\rm end}}^{2}}\frac{\cos\frac{\pi\beta}{2}}{(K|y_{\rm end}|)^\beta}
\Gamma(\beta+1) \left(\sum_i k_i^2\right)  \cdot \left\{\frac{K}{\beta-1} + \frac{1}{K}\sum_{i<j}k_ik_j +(1+\beta)\frac{k_1k_2k_3}{K^2}\right\}\,.
\label{gradzeta2}
\eea
\bea
\nonumber
{\cal A}^{(\bf II)}_{\dot \zeta \partial \zeta \partial \chi} &=& - \frac{\epsilon}{4 c_{s\ \rm end}^2 }\left\{\frac{\cos\frac{\pi\beta}{2} \Gamma(\beta+1)}{(K|y_{\rm end}|)^\beta}\left[\sum_j k_j^3 +\frac{\beta -1}{2} \sum_{i\neq j}k_i k_j^2 - 2(\beta+3) k_1 k_2 k_3 \right. \right.\\ \nonumber
&-& \left. \frac{2(\beta+1)}{K^2} \sum_{i\neq j} k_i^2 k_j^3 + \frac{3}{2 K k_1 k_2 k_3} \left(\sum_{i \neq j} k_i^2 k_j^5 - \sum_{i \neq j} k_i^3 k_j^4 \right)\right] \\ \nonumber
&+& 3 \left(\frac{\cos\frac{\pi\beta}{2} \Gamma(\beta)}{(K|y_{\rm end}|)^\beta} -\frac{1}{\beta}\right)
\left[-9 k_1 k_2 k_3 - 2 \sum_{i \neq j} k_i k_j^2 + \right. \\ 
&+& \left. \frac{1}{k_1 k_2 k_3} \left(\frac{1}{2} \sum_{i\neq j}k_i k_j^5
 -  \sum_{i< j}k_i^3 k_j^3\right)+   \frac{1}{2 k_1^2 k_2^2 k_3^2} \left(\sum_{i\neq j}k_i^3 k_j^6 - \sum_{i\neq j}k_i^4 k_j^5 \right)\right]\nonumber \\
&+& 3 K \left(\frac{\cos\frac{\pi\beta}{2} \Gamma(\beta-1)}{(K|y_{\rm end}|)^\beta} +\frac{1}{\beta}\right)\left[- 2 \sum_j k_j^2 - 9 \sum_{i< j}k_i k_j + \frac{1}{k_1^2 k_2^2 k_3^2}\left( \frac{1}{2} \sum_{i\neq j}k_i^2 k_j^6 - \sum_{i< j}k_i^4 k_j^4 \right)\right]\nonumber \\
&-& \left. 27\left[\frac{\cos\frac{\pi\beta}{2}}{(K|y_{\rm end}|)^\beta}\frac{\Gamma(\beta)}{\beta-2} + \frac{\beta-1}{\beta(\beta-2)}-\frac{1}{3}\right] \frac{K^3}{\beta-3} \right\}\, . \nonumber
\eea
\bea
\nonumber
{\cal A}^{(\bf II)}_{\epsilon^2} &=& \frac{\epsilon^2}{32 c_{s\ \rm end}^2 }\left\{\frac{\cos\frac{\pi\beta}{2} \Gamma(\beta+1)}{(K|y_{\rm end}|)^\beta}\left[(\beta+4) \left(\sum_j k_j^3 - \sum_{i\neq j}k_i k_j^2 + 2 k_1 k_2 k_3\right)\right. \right.\\ \nonumber
&-& \left.12 k_1 k_2 k_3 - \frac{3}{K} \sum_{i\neq j} k_i k_j^3 + \frac{3}{K k_1 k_2 k_3}(\sum_{i \neq j} k_i^2 k_j^5 - 2 \sum_{i \neq j} k_i^3 k_j^4 + \sum_{i \neq j} k_i k_j^6)\right] \\ \nonumber
&+& \left(\frac{\cos\frac{\pi\beta}{2} \Gamma(\beta)}{(K|y_{\rm end}|)^\beta} -\frac{1}{\beta}\right)
\left[-54 k_1 k_2 k_3 - 15 \sum_{i \neq j} k_i k_j^2 - 6 \sum_j k_j^3 + \frac{3}{k_1 k_2 k_3} \left(3 \sum_j k_j^6 \right. \right. \\
&-&\left. \left.  3 \sum_{i\neq j}k_i^2 k_j^4 + 2 \sum_{i\neq j}k_i k_j^5 - 4 \sum_{i< j}k_i^3 k_j^3\right)+   \frac{3}{k_1^2 k_2^2 k_3^2} \left(\sum_{i\neq j}k_i^3 k_j^6 
+ \sum_{i\neq j}k_i^2 k_j^7 - 2 \sum_{i\neq j}k_i^4 k_j^5\right)\right]\nonumber \\
&+& 3 K \left(\frac{\cos\frac{\pi\beta}{2} \Gamma(\beta-1)}{(K|y_{\rm end}|)^\beta} +\frac{1}{\beta}\right)\left[- 6 \sum_j k_j^2 - 18 \sum_{i< j}k_i k_j - \frac{3}{k_1 k_2 k_3}\left( \sum_{i\neq j}k_i^2 k_j^3 +\sum_{i\neq j}k_i k_j^4 \right. \right. \nonumber \\
&-&\left.\left. \sum_j k_k^5\right) + \frac{1}{k_1^2 k_2^2 k_3^2}\left( 2 \sum_{i\neq j}k_i^2 k_j^6 -4 \sum_{i< j}k_i^4 k_j^4 + 3 \sum_{i\neq j}k_i k_j^7 - 3 \sum_{i\neq j}k_i^3 k_j^5\right)\right] \nonumber \\
&-& \left. 9\left[\frac{\cos\frac{\pi\beta}{2}}{(K|y_{\rm end}|)^\beta}\frac{\Gamma(\beta)}{\beta-2} + \frac{\beta-1}{\beta(\beta-2)}-\frac{1}{3}\right] \frac{K^3}{\beta-3} \left[6 - \frac{1}{k_1^2 k_2^2 k_3^2} \left(\sum_j k_j^6 -  \sum_{i\neq j}k_i^2 k_j^4\right) \right] \right\} \nonumber \\
{\cal A}^{(\bf II)}_{\rm redef} &=&   -\frac{3(1+\epsilon)}{10c_{s\;{\rm end}}^2}\left[\sum_j k_j^3 \right]\,,
\label{3ampII}
\eea
where ${\cal A}^{(\bf II)}_{\rm redef}$ arises from the field redefinition~(\ref{fredef}) (\footnote{We are grateful to D. Baumann for pointing out important typos in an earlier version of this equation. The corrected expression appeared in~\cite{daniel}}).

Note that the expressions in Case I and II take nearly identical form in the limit of small sound speed, under the replacement $\alpha\leftrightarrow \beta$.
The only significant difference is the overall factor of ${\bar c}_s^{-2}$ in Case I compared to $c_{s\;{\rm end}}^2(|y_{\rm end}| K)^{\beta}$ in Case II, which
stems from the power spectrum being normalized at horizon-crossing and at the end of the contracting phase in Cases I and II, respectively.

\section{Non-Gaussian Signatures}
\label{ngsign}

In this Section, we discuss three observable features of the 3-point function, each of which can distinguish the expanding branch
from the contracting branch: the amplitude, parametrized by the $f_{\rm NL}$ parameter; the spectral dependence, characterized
 by the tilt $n_{\rm NG}$; and the momentum shape of the 3-point amplitude.

Because of the reduced sound speed the 3-point amplitude is relatively large and potentially measurable by
current or near-future CMB experiments. Interestingly, the $f_{\rm NL}$ parameter is generically negative in the expanding case
(including DBI inflation) and positive in the contracting case. In particular, this holds true for DBI lagrangians.
Nevertheless, it is certainly possible to construct models where this conclusion is altered.

In standard slow-roll inflation, including DBI, the quasi-de Sitter nature of the background ($\epsilon,\eta\ll 1$)
and the near constancy of the sound speed ($\epsilon_s\ll 1$) imply both a nearly scale-invariant power spectrum
as well as nearly scale invariant non-Gaussianities. In the generalized models studied here, however, the scale
invariance of the 2-point function is maintained by balancing a relatively large $\epsilon$ against a
correspondingly large $\epsilon_s$. This cancellation fails at the three-point level, leading to a strong scale
dependence for non-Gaussianities. Here again the expanding and contracting solutions generically
have a different signal: the inflationary branch (Case I) leads to a blue tilt for the 3-point amplitude,
whereas the ekpyrotic branch (Case II) has a red tilt.

But the most striking difference comes from the shape of the non-Gaussianities. In the inflationary branch, the shape is predominantly of the
equilateral type~\cite{paoloshape}, which is characteristic of inflationary models with low sound speed, as in DBI~\cite{DBI2}. In the ekpyrotic
branch, however, remarkably the shape is predominantly {\it local}. 

This stark contrast in shape arises from the different behavior of $\zeta$ on super-horizon scales. Because $\zeta$
tends to a constant outside the horizon in Case~I, non-linearities can only grow for a finite period until modes exit the horizon. 
Thus non-Gaussianities peak when all wavevectors have magnitude comparable to the Hubble radius, and therefore comparable to one another --- hence the
equilateral shape. In Case II, $\zeta$ keeps growing outside the horizon, thereby allowing non-linearities to grow until the end of the scaling phase. 
The non-Gaussian amplitude therefore peaks when one of the wavenumbers is small, corresponding to the squeezed-triangle limit.

There is also an interesting subdominant contribution to the non-Gaussianities. While this is suppressed by slow-roll parameters in standard DBI inflation,
here it accounts for a more significant fraction of the amplitude since $\epsilon$ can be relatively large in Case I. In the inflationary branch, we find that
the subdominant contribution is of the squashed shape; in the ekpyrotic branch, it is of the equilateral shape.

\subsection{Amplitude of Non-Gaussianities}

The 3-point amplitude is characterized as usual by the $f_{\rm NL}$ parameter~\cite{komatsuspergel}:
\be
\zeta = \zeta_g(x) + \frac{3}{5}f_{\rm NL}\zeta_g^2\,.
\ee
Here we follow the WMAP sign convention, where positive $f_{\rm NL}$ physically corresponds to negative-skewness for the temperature fluctuations.
The resulting 3-point function in momentum space is given by
\be
\langle \zeta(\textbf{k}_1)\zeta(\textbf{k}_2)\zeta(\textbf{k}_3)\rangle = (2\pi)^7 
\delta^3(\kk_1+\kk_2+\kk_3) P_\zeta^{\;2} \frac{3}{10}f_{\rm NL}\frac{\Sigma_j k_j^3}{\Pi_j k_j^3}\,,
\ee
which peaks in the squeezed limit, {\it e.g.}, $k_1 \ll k_2,k_3$. 

Although the momentum dependence of our amplitudes is manifestly different,
in particular peaking in the equilateral limit in Case I, it is conventional to define $f_{\rm NL}$ by matching amplitudes at  $k_1 = k_2 = k_3 = K/3$:
\be
f_{\rm NL} = 30\frac{{\cal A}_{k_1=k_2=k_3}}{K^3}\,.
\ee

\begin{figure}[htbp]
  \begin{center}
\includegraphics[width=3.2in]{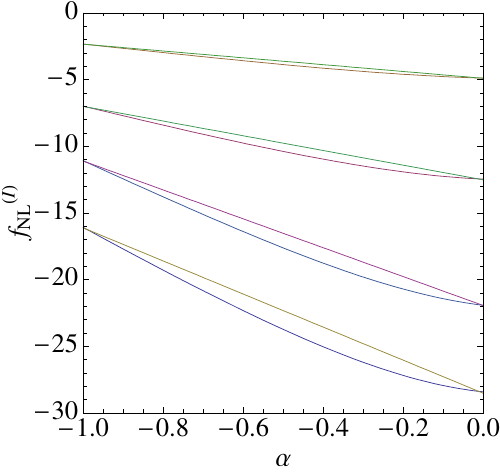}
  \end{center}
\caption{Comparison between the exact formula \eqref{FNLI} and its linear fitting \eqref{fNLfit1} in the expanding case, for the parameter region of interest: $-1<\alpha<0$. Values of $f_{NL}^{({\rm I})}$ are evaluated as a function of $\alpha$ for different $c_s$ and $f_X$: from top to bottom, $({\bar c}_s, f_X) = (0.25, 1);( 0.15,  0.1); (0.12,  0.9); (0.1, 0.1)$.} 
  \label{fit1a}
\end{figure}

In the expanding case, by evaluating the right-hand side for the amplitudes listed in Sec.~\ref{allterms}, we obtain
\bea 
\nonumber
f_{\rm NL}^{({\rm I})} &=& \frac{35}{108} - \frac{40}{243(\alpha+4)} \\ 
&-&\frac{10}{9\bar{c}_s^2(\alpha+4)}\cos\frac{\alpha\pi}{2}\Gamma(1+\alpha)\left\{\frac{2}{27}(\alpha+1)(\alpha+2)\left(f_X-1\right)+ \frac{7}{6}(1+\alpha)+\frac{9\alpha^2}{32}\right\}\,. 
\label{FNLI}
\eea
To gain some intuition on the parametric dependence, a useful fitting formula for this expression is
\be
f_{\rm NL}^{({\rm I})} \approx 0.27 - 0.164\, \bar{c}_s^{-2}- (0.12+0.04 f_X)\,  \bar{c}_s^{-2} (1+ \alpha)\,,
\label{fNLfit1}
\ee
where we recall from~(\ref{alpdef}) that $-2<\alpha < 0$ for $0<\epsilon<1$. Note that $f_{\rm NL}$ is fairly insensitive to
$f_X$, for reasonable values of this parameter, including DBI. Figure~\ref{fit1a} compares the fitting formula with the exact
expression as a function of $\alpha$ and for various values of $f_X$. The agreement is typically within 10 \% over the phenomenologically
allowed range of $\epsilon\lesssim 0.3$ (or $\alpha\; \gsim\; -1$).

\begin{figure}[htbp]
  \begin{center}
\includegraphics[width=3.2in]{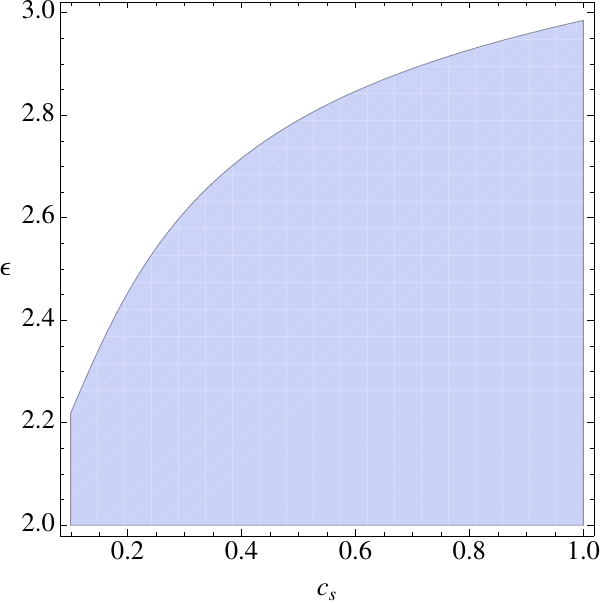}
 \end{center}
\caption{In theÊ $(c_{s\;{\rm end}}, \epsilon)$
plane we plot the region corresponding to $f_{\rm NL} < 100$ with
the choice $f_X=1$. The upper curve corresponds to the upper bound $f_{\rm
NL} = 100$.
}
\label{fit1}
\end{figure}

In the contracting case, on the other hand, the expression for $f_{\rm NL}$ is much more involved, as can be seen from~(\ref{3ampII}). Non-gaussianities are quite large for generic choice of parameters in this case. However acceptable values of $f_{\rm NL}$ can be obtained in some region of the parameter space. Figure~\ref{fit1} shows the region in the $(c_{s\;{\rm end}}, \epsilon)$ parameter space corresponding to $-100 < f_{\rm NL} < 100$, with $f_X=1$.

\subsection{Spectral dependence of $f_{\rm NL}$} \label{running}

Since the sound speed can vary rapidly in our class of models, the amplitude of non-Gaussianities will generically display a strong scale dependence.
Following~\cite{runningchen}, we define a spectral index for $f_{\rm NL}$ as
\be
n_{\rm NG} - 1 \equiv \frac{{\rm d}\ln f_{\rm NL}}{{\rm d}\ln K}\,.
\ee

In Case I, the non-Gaussian tilt arises from the $\bar{c}_s^{-2}$ prefactors in~(\ref{fNLfit1}). For small values of the sound speed, we obtain
\be
\left(n_{\rm NG} - 1\right)_{\rm I} \approx -2\frac{{\rm d}\ln \bar{c}_s}{{\rm d}\ln K} = -\alpha\,.
\ee
And since $-2<\alpha < 0$ within the inflationary range $0<\epsilon<1$, we get a blue tilt for the 3-point function in the expanding branch, corresponding to larger non-Gaussian amplitude on 
smaller scales. As discussed in Sec.~\ref{dura}, the present range of observable scales goes from the largest (CMB) ones down to typical galactic sizes: $K_{\rm gal}/K_{\rm CMB} \simeq 10^3$.
We therefore expect
\begin{equation}
f_{\rm NL}^{({\rm I})} ({\rm CMB})\ \approx \ 10^{3\alpha} f_{\rm NL}^{({\rm I})} ({\rm gal})\,.
\end{equation}
Clearly we must require that non-Gaussianities are within the perturbative regime down to galactic scales, \emph{i.e.}, $f_{\rm NL}^{({\rm I})} ({\rm gal}) \lesssim 10^4$. 
In the observationally-optimistic scenario in which large-scale non-Gaussianities are of the order of the present experimental limits ($|f_{\rm NL}({\rm CMB})|\sim 100$) this implies 
$|\alpha| \lesssim 2/3$ ($\epsilon \lesssim 1/5$). Our model is less constrained if we assume instead $|f_{\rm NL}({\rm CMB})| = {\cal O}(1)$, in which case $|\alpha | \lesssim 4/3$ ($\epsilon \lesssim 1/2$).

The tilt is also important in Case II. In this case, however, it cannot be approximated as constant over the relevant range of scales, since not all terms in~(\ref{3ampII}) share a common $(|y_{\rm end}| K)^{\beta}$ prefactor.

\begin{figure}[htbp]
  \begin{center}
     \subfigure[$-{\cal A}^{\rm (I)}(1,x_2,x_3)/(x_2 x_3)$ for $\alpha = -0.3$]{\label{fig2-a}\includegraphics[width=3in]{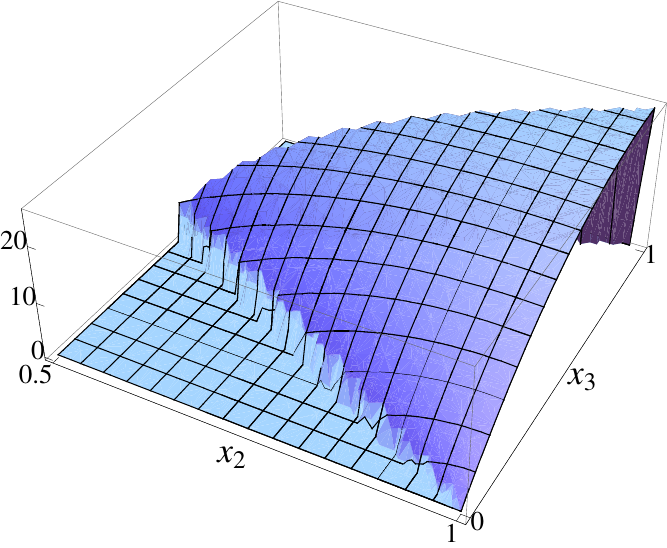}}
    \subfigure[$-{\cal A}^{\rm (I)}(1,x_2,x_3)/(x_2 x_3)$ for $\alpha = -0.9$]{\label{fig2-b}\includegraphics[width=3in]{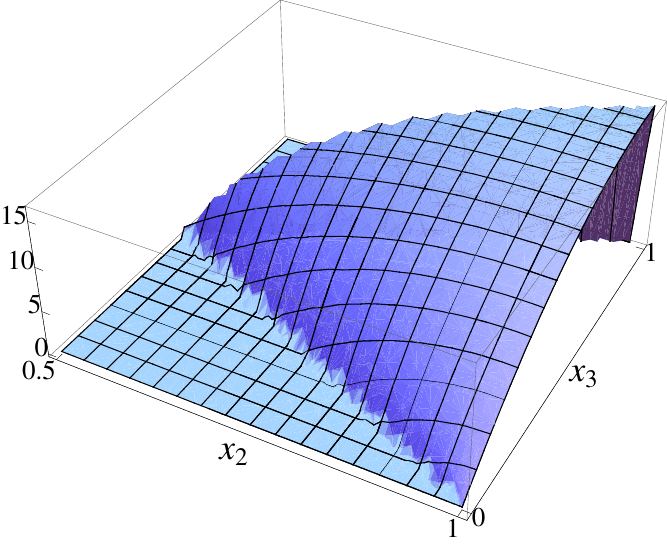}} \\
        \subfigure[$|{\cal A}^{\rm (I)} - A_{\rm equi}| /(x_2 x_3)$ for $\alpha = -0.3$]{\label{fig2-c}\includegraphics[width=3in]{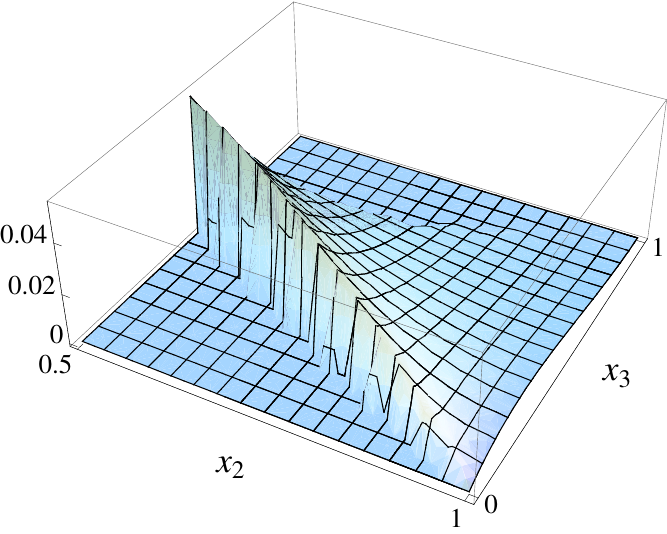}}
    \subfigure[$|{\cal A}^{\rm (I)} - A_{\rm equi}| /(x_2 x_3)$ for $\alpha = -0.9$]{\label{fig2-d}\includegraphics[width=3in]{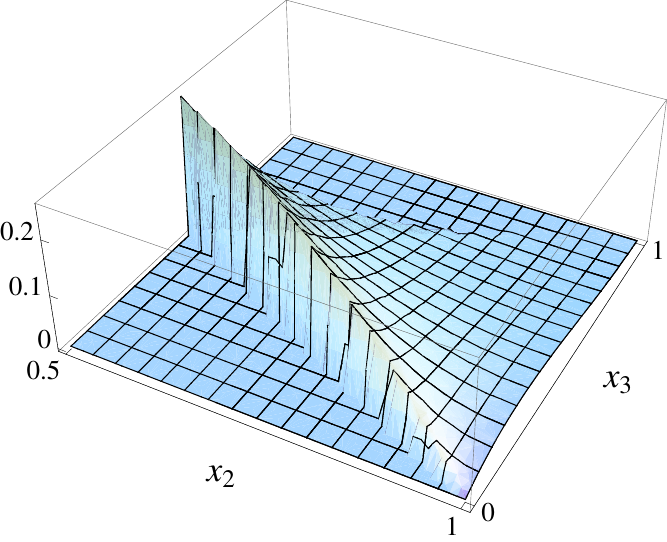}} \\
\end{center}
\caption{In the top two panels we plot the non-Gaussian ampitude in the inflationary branch, $-{\cal A}^{\rm (I)}(1,x_2,x_3)/(x_2 x_3)$ for ${\bar c}_{s} = 0.1$, $f_X=0.9$, $\alpha = -0.3$ ($\epsilon = 0.08$, top left) and $\alpha = -0.9$ ($\epsilon = 0.3$, top right). Thus the dominant contribution peaks for equilateral configurations ($x_2 \approx x_3 \approx 1$). To see how this differs from DBI inflation, in the bottom two panels we subtract the dominant DBI-like (equilateral) contribution and plot $|{\cal A}^{\rm (I)} - A_{\rm equi}| /(x_2 x_3)$, where ${\cal A}^{\rm (I)}$ and $A_{\rm equi}$ have each been normalized to one in the equilateral limit $x_2 = x_3 = 1$. Thus the subdominant contribution peaks for squashed configurations ($x_2 \approx x_3 \approx 0.5$).}
  \label{fig2}
\end{figure}

\subsection{Shape of non-Gaussianities}

In light of the substantial running of the three-point function, we study the shape of the amplitude {\it at fixed} $K$, thereby disentangling shape and running effects.
Following the literature~\cite{paoloshape}, it is convenient to focus on the dimensionless ratio ${\cal A}/k_1k_2k_3$.  We will plot this quantity 
${\cal A}/k_1k_2k_3$ in the $x_2$-$x_3$ plane, where $x_2=k_2/k_1$ and $x_3 = k_3/k_1$. Moreover, we can restrict ourselves to the range $1-x_2 \leq x_3 \leq x_2$,
where the first inequality is the triangle inequality while the second one avoids plotting the same configuration twice.

Starting with the expanding case, it is instructive to first consider the limit $\epsilon\rightarrow0$ ($\alpha\rightarrow0$). In this regime, 
the last two terms in~\eqref{3ampI} are subdominant and, save for the ${\cal A}_{\dot\zeta^3}^{(\rm I)}$ contribution, our shape reduces to the equilateral shape 
\begin{equation} \label{ADBI}
{\cal A}_{\rm equi} \ \propto \ \frac{1}{8}\sum_i k_i^3 - \frac{1}{K} \sum_{i<j} k_i^2 k_j^2 +\frac{1}{2 K^2} \sum_{i\neq j}k_i^2 k_j^3\, .
\end{equation} 
Now, even though each term above when divided by  $k_1k_2k_3$ is singular in the degenerate limit $k_3\rightarrow 0$, all divergences remarkably cancel each other in the combination~(\ref{ADBI}). One can in fact check that the cancellation works in the full set of amplitudes~\eqref{3ampI} for any value of  $\epsilon$ ($\alpha$), despite other potentially divergent terms such as $\sum_{i\neq j}k_i k_j^2$ and $\sum_i k_i^4$.
This is a consequence of the consistency relation \cite{malda} ${\cal A}/(k_1 k_2 k_3) \simeq (n_s - 1) k_1/k_3$ for the three-point function in the local limit $k_3 \ll k_1 = k_2$, which in this case predicts vanishing amplitude since $n_s = 1$. Therefore, the total amplitude 
peaks in the equilateral limit  $k_1 \approx k_2 \approx k_3$ for a generic choice of parameters, as shown in the top two panels of Fig.~\ref{fig2}, and the shape is predominantly of the equilateral type. This shape is generally expected in models with $c_s^2 <1$ and where higher derivatives operators are present in the effective action because of the shorter Jeans Length of the perturbations that easily produce non-linearities on shorter scales.

\begin{figure}[htbp] 
\begin{center} 
\includegraphics*[width=3.3in]{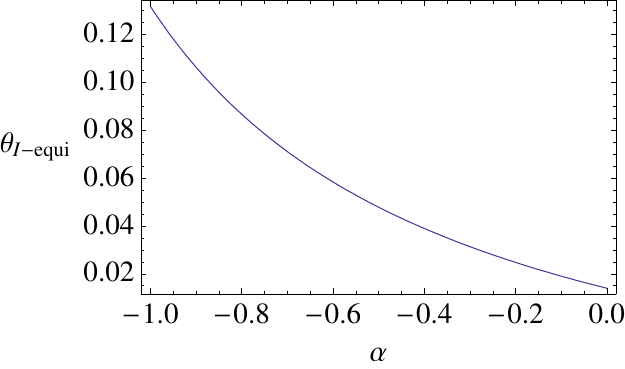}%
\hfil
\includegraphics*[width=3.3in]{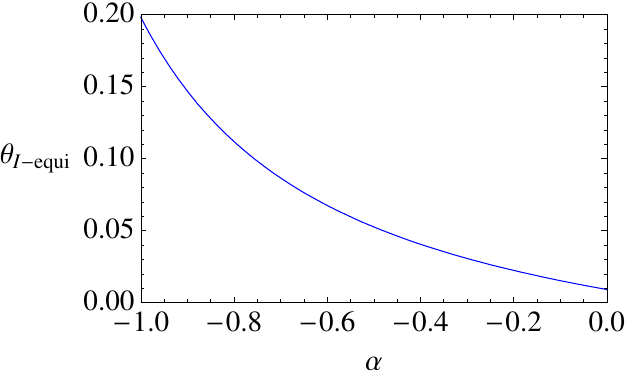}\\%
\end{center}\vspace{-.5cm}
\caption{The angle $\theta$ defined in eq. \eqref{theta} is plotted as a function of $\alpha$ in the expanding case. In the left panel ${\bar c}_s=0.05$ and $f_X =0.5$, in the right panel we choose instead ${\bar c}_s=0.5$ and $f_X =0.5$. The distinguishability between our shape and DBI's increases with $|\alpha |$ and with ${\bar c}_s$.}
\label{fig3} \end{figure}

Nevertheless, the shape in Case I is not exactly of the form~(\ref{ADBI}). We can unearth subdominant contributions by subtracting the equilateral shape~\eqref{ADBI}
normalized at the equilateral point $k_1 = k_2 = k_3$. The resulting amplitude, shown in the bottom two panels in Fig.~\ref{fig2}, peaks for ``squashed" triangles.
This subdominant contribution is of course also present in standard DBI inflation, albeit suppressed by slow roll parameters. Because our mechanism allows for
much larger values of $\epsilon$, the squashed amplitude is correspondingly more important here.

These statements can be made more quantitative by looking at the scalar product between shapes ${\cal A}_1 \cdot {\cal A}_2$, introduced by~\cite{paoloshape}
\be
{\cal A}_1 \cdot {\cal A}_2 = \sum_{\vec{k}_i}{\cal A}_1(k_1,k_2,k_3){\cal A}_2(k_1,k_2,k_3)/(\sigma_{k_1}^2\sigma_{k_2}^2\sigma_{k_3}^2)\,.
\ee
The corresponding angle $\theta_{12}$ therefore informs us on how different and distinguishable the two shapes are:
\begin{equation}\label{theta}
\cos(\theta_{12}) = \frac{{\cal A}_1 \cdot {\cal A}_2}{({\cal A}_1 \cdot {\cal A}_1)^{1/2}
({\cal A}_2 \cdot {\cal A}_2)^{1/2}}\,.
\end{equation}
In Fig.~\ref{fig3} we plot the angle above calculated between ${\cal A}^{\rm (I)}$ and ${\cal A}_{\rm equi}$ as a function of the parameter $\alpha$. As expected, the difference between the shapes increases with $|\alpha |$. 

\begin{figure}[htbp]
  \begin{center}
    \subfigure[$-{\cal A}^{\rm (II)}(1,x_2,x_3)/(x_2 x_3)$ for $\beta = 0.1$]{\label{fig4-a}\includegraphics[width=3in]{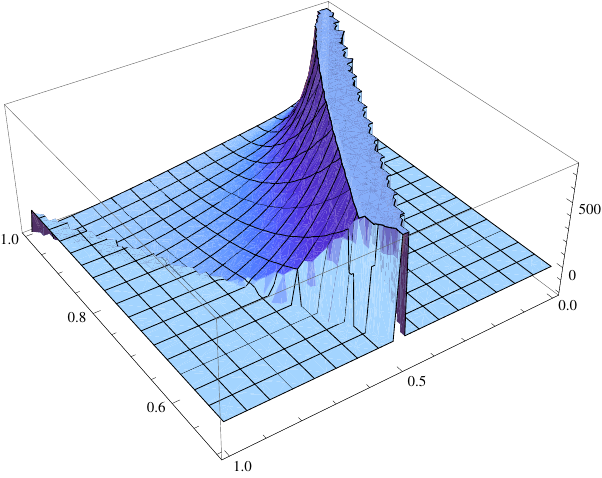}}
    \subfigure[$-{\cal A}^{\rm (II)}(1,x_2,x_3)/(x_2 x_3)$ for $\beta = -0.1$]{\label{fig4-b}\includegraphics[width=3in]{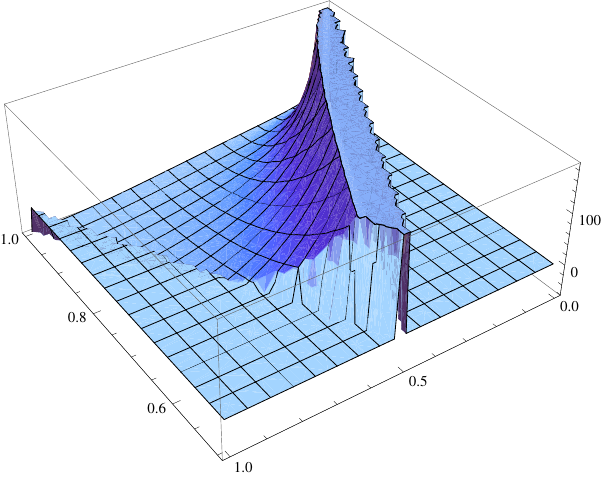}} 
    \end{center}
\caption{We plot ${\cal A}^{\rm (II)}(1,x_2,x_3)/(x_2 x_3)$ for ${c}_{s\, \rm end} = .95$, $f_X=1$, $\beta = 0.1$ (left) and ${c}_{s\, \rm end} = .9$,
$\beta = 0.01$ (right).}  \label{fig4}
\end{figure}

In the contracting case, as described earlier, the curvature perturbation keeps growing outside the horizon, resulting in non-Gaussianities that peak in the squeezed limit, as shown in Fig.~\ref{fig4}. Note, however, that the amplitude as $k_3\ll  k_1,k_2$, corresponding to squeezed configurations, is more singular than the local form:
\begin{equation}
{\cal A}_{\rm Loc} \propto \sum_i k_i^3 \,.
\label{ALoc}
\end{equation}
This traces back to higher powers of $k_i$ in the denominator of~(\ref{3ampII}). In particular, non-Gaussianity in the contracting case manifestly violates Maldacena's consistency relation, which
assumes that $\zeta(k_3)$ tends to a constant at long wavelengths and therefore behaves as a background wave for the short-wavelengths modes. This assumption does not hold in this case, while the reversed argument is probably the correct explanation for our shape a local type: precisely because $\zeta(k_3)$ keeps growing outside the horizon the short wavelength modes effectively ``feel" a different background evolution. 
The strong peak in the squeezed limit obtained here is remarkable since non-Gaussianity of this type usually results from entropy perturbations between two scalar degrees of freedom, such as in the curvaton~\cite{curvaton}, modulated reheating~\cite{DGZ}, pre-big bang pump field~\cite{pump} and New Ekpyrotic mechanisms~\cite{paolo,ngek,ngekwands,ngcyclic}.

\section{Conclusion}

The near scale invariance of the primordial spectrum estalished by CMB and large scale structure observations is widely hailed as evidence for a
phase of quasi de Sitter expansion in the early universe. In this paper we have shown that a suitably time-dependent sound speed allows
for a much richer spectrum of early universe cosmologies, including both expanding~\cite{cri} and contracting backgrounds.

In the expanding case, this mechanism was first proposed by~\cite{cri,joao,piao}, who considered a universe with an extremely large sound speed initially.
Unlike~\cite{joao} we have restricted ourselves to subluminal sound speed throughout. Thus the background must still be inflationary in order for
perturbations to be stretched outside the physical horizon. Inflationary expansion of course has the added virtue of making the universe flat, homogeneous and isotropic.
Our mechanism in this case alleviates the usual slow-roll conditions, from the restrictive $\epsilon \lesssim 1/60$ to $\epsilon \lesssim 0.3$.

In the contracting branch, requiring that the background addresses the flatness and homogeneity problems of standard big bang cosmology leads to the ekpyrotic regime
$\epsilon > 3$ (or $w > 1$). Unlike the New Ekpyrotic mechanism which relies on two scalar fields, here a single scalar degree of freedom is sufficient to endow $\zeta$ with
a scale invariant spectrum.

The phenomenology of non-Gaussianities is very rich and offers many distinguishing features. To start with, the cancelation mechanism 
between the sound speed and the cosmological equation of state does not carry over to the three-point function. Hence non-Gaussianities
display a strong scale dependence, peaking on small scales in the inflationary branch and on large scales in the ekpyrotic branch. 

Moreover, the shape of the amplitude is drastically different in each case. In the expanding case, the amplitude is predominantly of the equilateral type,
as usual with low sound speed models, such as DBI inflation. In the contracting case, it is predominantly of the local form, peaking for squeezed triangles.
This contrast in shape arises because $\zeta$ has different behaviors on super-horizon scales: whereas it tends to a constant as usual in the inflationary case,
it keeps growing in the ekpyrotic branch. Beneath the dominant shape lies a subdominant contribution, which can make up as much as 10\% of the total amplitude. 
This subdominant piece peaks for squashed and equilateral triangles in the inflationary and ekpyrotic branches, respectively.

Many offspring projects naturally suggest themselves. Firstly, while our study of the non-Gaussianity focused on $P(X,\phi)$ scalar lagrangians, it would
be interesting to consider the more general effective field theory framework proposed in~\cite{effinf}. From this point of view, the $P(X,\phi)$ predictions
are a subset of all possible results --- hence we expect a wider spectrum of non-Gaussian features. Secondly, in light of the abundance of non-Gaussian features in our class of
models, it will be enlightning to investigate the potential of near-future experiments such as Planck to detect some of the signatures described here, such as the strong title
and the distinctive shapes. Moreover, we should push the non-Gaussian exploration further and calculate higher-point correlation functions in this framework.
The shape and tilt of the trispectrum should offer more ways to break degeneracies between the different cosmologies studied here.

{\bf Acknowledgments}
We thank Niayesh~Afshordi, Robert Brandenberger, Maurizio Gasperini, Jean-Luc~Lehners, Alberto~Nicolis, Leonardo Senatore, Paul~Steinhardt,  Gabriele Veneziano, Matias Zaldarriaga and especially Andrew~Tolley for many helpful discussions. We are grateful to Daniel Baumann for helpful feedback and for pointing out important typos in the original version of Eq. \eqref{gradzeta2}. The research of J.K. and F.P. at the Perimeter Institute is supported in part by the Government of Canada through NSERC and by the Province of Ontario through
the Ministry of Research \& Innovation.\\

\appendix

\section{Explicit DBI Examples} \label{appendixA}

In this Section, we present explicit realizations of the scaling solutions described so far, closely following the analysis of~\cite{kinney}. We focus on DBI lagrangians:
\be
{\cal L}_{\rm DBI} =  -f^{-1}(\phi)\sqrt{1+f(\phi)(\partial\phi)^2}+f^{-1}(\phi)-V(\phi)\,,
\ee
where $f(\phi)$ describes the warp factor of the throat within which the 3-brane is moving. The cosmological equations are
\bea
\nonumber
3H^2M_{\rm Pl}^2 &=& \frac{c_s^{-1} -1}{f} + V\;; \\
\frac{{\rm d}\ln H}{{\rm d}N} &=& -\frac{1}{2M_{\rm Pl}^2c_s}\left(\frac{{\rm d}\phi}{{\rm d}N}\right)^2\,,
\label{eomDBI}
\eea
where ${\rm d}N\equiv H{\rm d}t$, and $c_s$ is given in~(\ref{DBIquantities}). 

Our goal is to determine the form of $f(\phi)$ and $V(\phi)$ corresponding to the scaling solutions of interest. 
To begin with, along the solution $c_s\sim e^{\epsilon_sN}$, it is useful to think of $c_s$ as a function of the classical solution for $\phi$.
Thus,
\be
\epsilon_s = \frac{{\rm d}\ln c_s(\phi)}{{\rm d}\phi}\frac{{\rm d}\phi}{{\rm d}N} = \pm \frac{\sqrt{2\epsilon}}{c_s^{1/2}}\frac{{\rm d}c_s}{{\rm d}\phi}M_{\rm Pl}\,,
\label{csode}
\ee
where in the last step we have used~(\ref{eomDBI}). And since $\epsilon_s$ is assumed constant, this can be integrated at once for $c_s(\phi)$:
\be
c_s(\phi) = \frac{\epsilon_s^2}{8\epsilon}\frac{\phi^2}{M_{\rm Pl}^2}\,.
\ee
Note that we have chosen the integration constant such that the sound speed vanishes when $\phi = 0$.

Similarly, on the background solution we can think of $H$ as a function of $\phi(N)$. Proceeding along similar lines as~(\ref{csode}), $\epsilon$ can be expressed as
\be
\epsilon = 2M_{\rm Pl}^2c_s\left(\frac{{\rm d}\ln H}{{\rm d}\phi}\right)^2\,.
\ee
And substituting for $c_s(\phi)$ yields a differential equation for the Hubble parameter: ${\rm d}\ln H/{\rm d}\ln \phi = \pm 2\epsilon/\epsilon_s$, which easily integrates to
\be
H(\phi) \sim \phi^{-2\epsilon/\epsilon_s}\,.
\ee
Noting that $\epsilon_s < 0$ in both Cases I and II --- in Case II, this assumes $\epsilon > 3/2$ ---, we have chosen the sign so that $|H|$ decreases (increases)
in the expanding (contracting) case. 

Having obtained expressions for $c_s(\phi)$ and $H(\phi)$, we can solve~(\ref{eomDBI}) for the potential and warp factor:
\be
V(\phi) = 3H^2(\phi)M_{\rm Pl}^2\left(1-\frac{2\epsilon}{3(1+c_s(\phi))}\right)\,.
\ee
Explicitly, this gives
\be
V(\phi) = V_0\left(\frac{\phi}{M_{\rm Pl}}\right)^{-4\epsilon/\epsilon_s}\left(1- \frac{2\epsilon}{3}\frac{1}{1+ \frac{\epsilon_s^2\phi^2}{8\epsilon M_{\rm Pl}^2}}\right)\,,
\ee
where $V_0$ is a constant. For Case I, where $\epsilon_s = -2\epsilon$, and in the limit of small sound speed ($\phi\ll M_{\rm Pl}$), this gives
\be
V_{\rm I}(\phi) \approx V_0\left(1-\frac{2\epsilon}{3}\right)\frac{\phi^2}{M_{\rm Pl}^2}\,.
\label{VI}
\ee
Meanwhile, for Case II, where $\epsilon_s = 2(3-2\epsilon)/5$,
\be
V_{\rm II}(\phi) \approx -V_0\left(\frac{2\epsilon}{3}-1\right)\left(\frac{\phi}{M_{\rm Pl}}\right)^{10\epsilon/(2\epsilon-3)}\,.
\ee 

Similarly, the warp factor can be expressed as
\be
f(\phi) = \frac{1-c_s^2(\phi)}{2\epsilon H^2(\phi)M_{\rm Pl}^2c_s(\phi)}\,,
\ee
which, for our explicit solutions, reduces to 
\be
f(\phi) = \frac{12}{V_0\epsilon_s^2}\left(\frac{\phi}{M_{\rm Pl}}\right)^{\frac{4\epsilon}{\epsilon_s}-2}\left(1-\frac{\epsilon_s^4\phi^4}{64\epsilon^2M_{\rm Pl}^4}\right)\,.
\ee
Again in the limit of small sound speed, we obtain, respectively for Case I and II,
\bea
\nonumber
f_{\rm I}(\phi) &\sim& \left(\frac{\phi}{M_{\rm Pl}}\right)^{-4}\,; \\
f_{\rm II}(\phi) &\sim&  \left(\frac{\phi}{M_{\rm Pl}}\right)^{-2\frac{7\epsilon-3}{2\epsilon-3}}\,.
\eea

Thus the desired potentials and warp factors are approximately power-law in form. Note that in Case I we have reproduced 
 the potential and AdS warp factor of~\cite{DBI2}. The exponents in $V(\phi)$ and $f(\phi)$ are independent of $\epsilon$ in this case.

\section{Relevant Integrals}

\label{appendixB}

Below are the integrals used in the calculations of Section \ref{7} for Case II.
\bea
& & {\rm Im}\left\{ \int_{-\infty}^{y_{\rm end}}{\rm d}y\frac{H}{c_s^{5/2}}\frac{1-iKy}{y^4}e^{iKy}\right\} = -\frac{H_{\rm end}}{c_{s\;{\rm end}}^{5/2}}\frac{K^3}{\beta-3}\left[\frac{\cos\frac{\pi\beta}{2}}{(K|y_{\rm end}|)^\beta}\frac{\Gamma(\beta)}{\beta-2} + \frac{\beta-1}{\beta(\beta-2)}-\frac{1}{3}\right] \\
& & {\rm Im}\left\{ \int_{-\infty}^{y_{\rm end}}{\rm d}y\frac{H}{c_s^{1/2}}\frac{1-iKy}{y^4}e^{iKy}\right\} = -\frac{H_{\rm end}}{c_{s\;{\rm end}}^{1/2}}\frac{K^3}{9} \\
& & {\rm Im}\left\{ \int_{-\infty}^{y_{\rm end}}{\rm d}y\frac{H}{c_s^{5/2}}\frac{e^{iKy}}{y^2}\right\} = \frac{H_{\rm end}}{c_{s\;{\rm end}}^{5/2}}K\left[\frac{\cos\frac{\pi\beta}{2}}{(K|y_{\rm end}|)^\beta}\Gamma(\beta-1) +\frac{1}{\beta}\right] \\
& & {\rm Im}\left\{ \int_{-\infty}^{y_{\rm end}}{\rm d}y\frac{H}{c_s^{1/2}}\frac{e^{iKy}}{y^2}\right\} = -\frac{H_{\rm end}}{c_{s\;{\rm end}}^{1/2}}\frac{K}{3} \\
& & {\rm Im}\left\{ i \int_{-\infty}^{y_{\rm end}}{\rm d}y\frac{H}{c_s^{5/2}}\frac{e^{iKy}}{y}\right\} = -\frac{H_{\rm end}}{c_{s\;{\rm end}}^{5/2}}\left[\frac{\cos\frac{\pi\beta}{2}}{(K|y_{\rm end}|)^\beta}\Gamma(\beta)-\frac{1}{\beta}\right]  \\
& & {\rm Im}\left\{ i \int_{-\infty}^{y_{\rm end}}{\rm d}y\frac{H}{c_s^{1/2}}\frac{e^{iKy}}{y}\right\} = -\frac{1}{3}\frac{H_{\rm end}}{c_{s\;{\rm end}}^{1/2}} \\
& & {\rm Im}\left\{\int_{-\infty}^{y_{\rm end}}{\rm d}y\frac{H}{c_s^{5/2}}e^{iKy}\right\} = -\frac{H_{\rm end}}{c_{s\;{\rm end}}^{5/2}}\frac{1}{K}\frac{\cos\frac{\pi\beta}{2}}{(K|y_{\rm end}|)^\beta}\Gamma(\beta+1) \\
& & {\rm Im}\left\{\int_{-\infty}^{y_{\rm end}}{\rm d}y\frac{H}{c_s^{1/2}}e^{iKy}\right\} \sim (Ky_{\rm end})^2 \Longrightarrow {\rm subleading} \\
& & {\rm Im}\left\{ i \int_{-\infty}^{y_{\rm end}}{\rm d}y\frac{H}{c_s^{5/2}}y e^{iKy}\right\}  = \frac{H_{\rm end}}{c_{s\;{\rm end}}^{5/2}}\frac{1}{K^2}\frac{\cos\frac{\pi\beta}{2}}{(K|y_{\rm end}|)^\beta}\Gamma(\beta+2) \\
& & {\rm Im}\left\{ i \int_{-\infty}^{y_{\rm end}}{\rm d}y\frac{H}{c_s^{1/2}}y e^{iKy}\right\}\sim (Ky_{\rm end})^2 \Longrightarrow {\rm subleading} \\
& & {\rm Im}\left\{\int_{-\infty}^{y_{\rm end}}{\rm d}y\frac{H}{c_s^{5/2}}y^2 e^{iKy}\right\}  = \frac{H_{\rm end}}{c_{s\;{\rm end}}^{5/2}}\frac{1}{K^3}\frac{\cos\frac{\pi\beta}{2}}{(K|y_{\rm end}|)^\beta}\Gamma(\beta+3) \\
& & {\rm Im}\left\{\int_{-\infty}^{y_{\rm end}}{\rm d}y\frac{H}{c_s^{1/2}}y^2 e^{iKy}\right\}  \sim (Ky_{\rm end})^2 \Longrightarrow {\rm subleading} \\
& & {\rm Im}\left\{i \int_{-\infty}^{y_{\rm end}}{\rm d}y\frac{H}{c_s^{5/2}}y^3 e^{iKy}\right\}  = - \frac{H_{\rm end}}{c_{s\;{\rm end}}^{5/2}}\frac{1}{K^4}\frac{\cos\frac{\pi\beta}{2}}{(K|y_{\rm end}|)^\beta}\Gamma(\beta+4)\,.
\eea




\begin{thebibliography}{99}


\bibitem{dust}
  F.~Finelli and R.~Brandenberger,
  ``On the generation of a scale-invariant spectrum of adiabatic  fluctuations
  in cosmological models with a contracting phase,''
  Phys.\ Rev.\  D {\bf 65}, 103522 (2002)
  [arXiv:hep-th/0112249];
    D.~Wands,
  ``Duality invariance of cosmological perturbation spectra,''
  Phys.\ Rev.\  D {\bf 60}, 023507 (1999)
  [arXiv:gr-qc/9809062].

\bibitem{gratton}
    S.~Gratton, J.~Khoury, P.~J.~Steinhardt and N.~Turok,
  ``Conditions for generating scale-invariant density perturbations,''
  Phys.\ Rev.\  D {\bf 69}, 103505 (2004)
  [arXiv:astro-ph/0301395].


\bibitem{newek} 
  E.~I.~Buchbinder, J.~Khoury and B.~A.~Ovrut,
  ``New Ekpyrotic Cosmology,''
  Phys.\ Rev.\  D {\bf 76}, 123503 (2007)
  [arXiv:hep-th/0702154].


\bibitem{fabio}
  F.~Finelli,
  ``Assisted contraction,''
  Phys.\ Lett.\  B {\bf 545}, 1 (2002)
  [arXiv:hep-th/0206112];
   S.~Tsujikawa, R.~Brandenberger and F.~Finelli,
  ``On the construction of nonsingular pre-big-bang and ekpyrotic cosmologies
  and the resulting density perturbations,''
  Phys.\ Rev.\  D {\bf 66}, 083513 (2002)
  [arXiv:hep-th/0207228].

\bibitem{lehners}
  J.~L.~Lehners, P.~McFadden, N.~Turok and P.~J.~Steinhardt,
  ``Generating ekpyrotic curvature perturbations before the big bang,''
  Phys.\ Rev.\  D {\bf 76}, 103501 (2007)
  [arXiv:hep-th/0702153].

\bibitem{paolo}
  P.~Creminelli and L.~Senatore,
  ``A smooth bouncing cosmology with scale invariant spectrum,''
  JCAP {\bf 0711}, 010 (2007)
  [arXiv:hep-th/0702165].

\bibitem{wands}
  K.~Koyama and D.~Wands,
  ``Ekpyrotic collapse with multiple fields,''
  JCAP {\bf 0704}, 008 (2007)
  [arXiv:hep-th/0703040];
    K.~Koyama, S.~Mizuno and D.~Wands,
  ``Curvature perturbations from ekpyrotic collapse with multiple fields,''
  Class.\ Quant.\ Grav.\  {\bf 24}, 3919 (2007)
  [arXiv:0704.1152 [hep-th]].

\bibitem{notari}
  A.~Notari and A.~Riotto,
  ``Isocurvature perturbations in the ekpyrotic universe,''
  Nucl.\ Phys.\  B {\bf 644}, 371 (2002)
  [arXiv:hep-th/0205019].

\bibitem{pbb}
 G.~Veneziano,
  ``Scale Factor Duality For Classical And Quantum Strings,''
  Phys.\ Lett.\  B {\bf 265}, 287 (1991);
  M.~Gasperini and G.~Veneziano,
  ``Pre - big bang in string cosmology,''
  Astropart.\ Phys.\  {\bf 1}, 317 (1993)
  [arXiv:hep-th/9211021].

\bibitem{pump}
  E.~J.~Copeland, R.~Easther and D.~Wands,
  ``Vacuum fluctuations in axion-dilaton cosmologies,''
  Phys.\ Rev.\  D {\bf 56}, 874 (1997)
  [arXiv:hep-th/9701082];
    E.~J.~Copeland, J.~E.~Lidsey and D.~Wands,
  ``S-duality-invariant perturbations in string cosmology,''
  Nucl.\ Phys.\  B {\bf 506}, 407 (1997)
  [arXiv:hep-th/9705050];
    E.~J.~Copeland, A.~Lahiri and D.~Wands,
  ``Low-energy effective string cosmology,''
  Phys.\ Rev.\  D {\bf 50}, 4868 (1994)
  [arXiv:hep-th/9406216];
    E.~J.~Copeland, A.~Lahiri and D.~Wands,
  ``String cosmology with a time dependent antisymmetric tensor potential,''
  Phys.\ Rev.\  D {\bf 51}, 1569 (1995)
  [arXiv:hep-th/9410136];
   M.~Gasperini and G.~Veneziano,
  ``Constraints on pre-big bang models for seeding large-scale anisotropy  by
  massive Kalb-Ramond axions,''
  Phys.\ Rev.\  D {\bf 59}, 043503 (1999)
  [arXiv:hep-ph/9806327].

 \bibitem{jlrev}
   J.~L.~Lehners,
  ``Ekpyrotic and Cyclic Cosmology,''
  Phys.\ Rept.\  {\bf 465}, 223 (2008)
  [arXiv:0806.1245 [astro-ph]].

\bibitem{pbbrev}
  M.~Gasperini and G.~Veneziano,
  ``The pre-big bang scenario in string cosmology,''
  Phys.\ Rept.\  {\bf 373}, 1 (2003)
  [arXiv:hep-th/0207130].

\bibitem{mollerach}
S.~Mollerach,
  ``Isocurvature Baryon Perturbations And Inflation,''
  Phys.\ Rev.\  D {\bf 42}, 313 (1990).



\bibitem{curvaton}
 D.~H.~Lyth and D.~Wands,
  ``Generating the curvature perturbation without an inflaton,''
  Phys.\ Lett.\  B {\bf 524}, 5 (2002)
  [arXiv:hep-ph/0110002];
    K.~Enqvist and M.~S.~Sloth,
  ``Adiabatic CMB perturbations in pre big bang string cosmology,''
  Nucl.\ Phys.\  B {\bf 626}, 395 (2002)
  [arXiv:hep-ph/0109214];
    T.~Moroi and T.~Takahashi,
  ``Effects of cosmological moduli fields on cosmic microwave background,''
  Phys.\ Lett.\  B {\bf 522}, 215 (2001)
  [Erratum-ibid.\  B {\bf 539}, 303 (2002)]
  [arXiv:hep-ph/0110096].

\bibitem{pbbcurvaton}
V.~Bozza, M.~Gasperini, M.~Giovannini and G.~Veneziano,
   ``Assisting pre-big bang phenomenology through short-lived axions,''
   Phys.\ Lett.\  B {\bf 543}, 14 (2002)
   [arXiv:hep-ph/0206131];
V.~Bozza, M.~Gasperini, M.~Giovannini and G.~Veneziano,
   ``Constraints on pre-big bang parameter space from CMBR anisotropies,''
   Phys.\ Rev.\  D {\bf 67}, 063514 (2003)
   [arXiv:hep-ph/0212112].

  \bibitem{maurizio}
  M.~Gasperini,
  ``Tensor perturbations in high-curvature string backgrounds,''
  Phys.\ Rev.\  D {\bf 56}, 4815 (1997)
  [arXiv:gr-qc/9704045].
  
\bibitem{cri}
  C.~Armendariz-Picon and E.~A.~Lim,
  ``Scale invariance without inflation?,''
  JCAP {\bf 0312}, 002 (2003)
  [arXiv:astro-ph/0307101]; 
  C.~Armendariz-Picon,
  ``Near scale invariance with modified dispersion relations,''
  JCAP {\bf 0610}, 010 (2006)
  [arXiv:astro-ph/0606168].
  
  \bibitem{joao}
  J.~Magueijo,
  ``Speedy sound and cosmic structure,''
  Phys.\ Rev.\ Lett.\  {\bf 100}, 231302 (2008)
  [arXiv:0803.0859 [astro-ph]].
  
  \bibitem{piao}
  Y.~S.~Piao,
  ``Seeding of Primordial Perturbations During a Decelerated Expansion,''
  Phys.\ Rev.\  D {\bf 75}, 063517 (2007)
  [arXiv:gr-qc/0609071].

\bibitem{IRUV}
  A.~Adams, N.~Arkani-Hamed, S.~Dubovsky, A.~Nicolis and R.~Rattazzi,
  ``Causality, analyticity and an IR obstruction to UV completion,''
  JHEP {\bf 0610}, 014 (2006)
  [arXiv:hep-th/0602178].

\bibitem{runningchen}
  X.~Chen,
  ``Running non-Gaussianities in DBI inflation,''
  Phys.\ Rev.\  D {\bf 72}, 123518 (2005)
  [arXiv:astro-ph/0507053].

\bibitem{loverde}
  M.~LoVerde, A.~Miller, S.~Shandera and L.~Verde,
  ``Effects of Scale-Dependent Non-Gaussianity on Cosmological Structures,''
  JCAP {\bf 0804}, 014 (2008)
  [arXiv:0711.4126 [astro-ph]].


\bibitem{paoloshape}
  D.~Babich, P.~Creminelli and M.~Zaldarriaga,
  ``The shape of non-Gaussianities,''
  JCAP {\bf 0408}, 009 (2004)
  [arXiv:astro-ph/0405356].

\bibitem{eva}
  E.~Silverstein and D.~Tong,
  ``Scalar speed limits and cosmology: Acceleration from D-cceleration,''
  Phys.\ Rev.\  D {\bf 70}, 103505 (2004)
  [arXiv:hep-th/0310221].

\bibitem{DBI2}
M.~Alishahiha, E.~Silverstein and D.~Tong,
``DBI in the sky,''
Phys.\ Rev.\ D {\bf 70}, 123505 (2004)
[arXiv:hep-th/0404084].

\bibitem{chen}
  X.~Chen, M.~x.~Huang, S.~Kachru and G.~Shiu,
  ``Observational signatures and non-Gaussianities of general single field
  inflation,''
  JCAP {\bf 0701}, 002 (2007)
  [arXiv:hep-th/0605045].

\bibitem{komatsuspergel}
 E.~Komatsu and D.~N.~Spergel,
  ``Acoustic signatures in the primary microwave background bispectrum,''
  Phys.\ Rev.\  D {\bf 63}, 063002 (2001)
  [arXiv:astro-ph/0005036].


\bibitem{DGZ}
  G.~Dvali, A.~Gruzinov and M.~Zaldarriaga,
  ``A new mechanism for generating density perturbations from inflation,''
  Phys.\ Rev.\  D {\bf 69}, 023505 (2004)
  [arXiv:astro-ph/0303591];
  L.~Kofman,
  ``Probing string theory with modulated cosmological fluctuations,''
  arXiv:astro-ph/0303614.

\bibitem{ngek}
  E.~I.~Buchbinder, J.~Khoury and B.~A.~Ovrut,
  ``Non-Gaussianities in New Ekpyrotic Cosmology,''
  Phys.\ Rev.\ Lett.\  {\bf 100}, 171302 (2008)
  [arXiv:0710.5172 [hep-th]].

\bibitem{ngekwands}
 K.~Koyama, S.~Mizuno, F.~Vernizzi and D.~Wands,
  ``Non-Gaussianities from ekpyrotic collapse with multiple fields,''
  JCAP {\bf 0711}, 024 (2007)
  [arXiv:0708.4321 [hep-th]].

\bibitem{ngcyclic}
 J.~L.~Lehners and P.~J.~Steinhardt,
  ``Non-Gaussian Density Fluctuations from Entropically Generated Curvature
  Perturbations in Ekpyrotic Models,''
  Phys.\ Rev.\  D {\bf 77}, 063533 (2008)
  [arXiv:0712.3779 [hep-th]];
  ``Intuitive understanding of non-gaussianity in ekpyrotic and cyclic
  models,''
  Phys.\ Rev.\  D {\bf 78}, 023506 (2008)
  [arXiv:0804.1293 [hep-th]].


\bibitem{bondsalopek}
  D.~S.~Salopek and J.~R.~Bond,
  ``Stochastic inflation and nonlinear gravity,''
  Phys.\ Rev.\  D {\bf 43}, 1005 (1991).
  
  \bibitem{malda}
  J.~M.~Maldacena,
  ``Non-Gaussian features of primordial fluctuations in single field
  inflationary models,''
  JHEP {\bf 0305}, 013 (2003)
  [arXiv:astro-ph/0210603].

\bibitem{bst}
  J.~M.~Bardeen, P.~J.~Steinhardt and M.~S.~Turner,
  ``Spontaneous Creation Of Almost Scale - Free Density Perturbations In An
  Inflationary Universe,''
  Phys.\ Rev.\  D {\bf 28}, 679 (1983).

\bibitem{proof}
  D.~H.~Lyth, K.~A.~Malik and M.~Sasaki,
  ``A general proof of the conservation of the curvature perturbation,''
  JCAP {\bf 0505}, 004 (2005)
  [arXiv:astro-ph/0411220].

\bibitem{tolleywesley}
  A.~J.~Tolley and D.~H.~Wesley,
  ``Scale-invariance in expanding and contracting universes from two-field
  models,''
  JCAP {\bf 0705}, 006 (2007)
  [arXiv:hep-th/0703101].

\bibitem{newekinit}
    E.~I.~Buchbinder, J.~Khoury and B.~A.~Ovrut,
  ``On the Initial Conditions in New Ekpyrotic Cosmology,''
  JHEP {\bf 0711}, 076 (2007)
  [arXiv:0706.3903 [hep-th]].

\bibitem{mixmaster}
 J.~K.~Erickson, D.~H.~Wesley, P.~J.~Steinhardt and N.~Turok,
  ``Kasner and mixmaster behavior in universes with equation of state 
 $w \geq 1$,''
  Phys.\ Rev.\  D {\bf 69}, 063514 (2004)
  [arXiv:hep-th/0312009].

\bibitem{kinney}
  W.~H.~Kinney and K.~Tzirakis,
  ``Quantum modes in DBI inflation: exact solutions and constraints from vacuum
  selection,''
  Phys.\ Rev.\  D {\bf 77}, 103517 (2008)
  [arXiv:0712.2043 [astro-ph]].

\bibitem{garrigamukhanov}
  J.~Garriga and V.~F.~Mukhanov,
  ``Perturbations in k-inflation,''
  Phys.\ Lett.\  B {\bf 458}, 219 (1999)
  [arXiv:hep-th/9904176].

\bibitem{BKL}
  E.~M.~Lifshitz and I.~M.~Khalatnikov,
  ``Investigations in relativistic cosmology,''
  Adv.\ Phys.\  {\bf 12}, 185 (1963);
  V.~A.~Belinsky, I.~M.~Khalatnikov and E.~M.~Lifshitz,
  ``Oscillatory approach to a singular point in the relativistic cosmology,''
  Adv.\ Phys.\  {\bf 19}, 525 (1970).
  
\bibitem{chaosothers}
 J.~Demaret, M.~Henneaux and P.~Spindel,
  ``Nonoscillatory Behavior In Vacuum Kaluza-Klein Cosmologies,''
  Phys.\ Lett.\  B {\bf 164}, 27 (1985); 
   T.~Damour and M.~Henneaux,
  ``Chaos in superstring cosmology,''
  Phys.\ Rev.\ Lett.\  {\bf 85}, 920 (2000)
  [arXiv:hep-th/0003139];
   T.~Damour and M.~Henneaux,
  ``E(10), BE(10) and arithmetical chaos in superstring cosmology,''
  Phys.\ Rev.\ Lett.\  {\bf 86}, 4749 (2001)
  [arXiv:hep-th/0012172];
     T.~Damour, M.~Henneaux and H.~Nicolai,
  ``Cosmological billiards,''
  Class.\ Quant.\ Grav.\  {\bf 20}, R145 (2003)
  [arXiv:hep-th/0212256].

\bibitem{ek1}
 J.~Khoury, B.~A.~Ovrut, P.~J.~Steinhardt and N.~Turok,
  ``The ekpyrotic universe: Colliding branes and the origin of the hot big
  bang,''
  Phys.\ Rev.\  D {\bf 64}, 123522 (2001)
  [arXiv:hep-th/0103239].

\bibitem{ek2}  
    J.~Khoury, B.~A.~Ovrut, N.~Seiberg, P.~J.~Steinhardt and N.~Turok,
  ``From big crunch to big bang,''
  Phys.\ Rev.\  D {\bf 65}, 086007 (2002)
  [arXiv:hep-th/0108187].

\bibitem{ek3}
  J.~Khoury, B.~A.~Ovrut, P.~J.~Steinhardt and N.~Turok,
  ``Density perturbations in the ekpyrotic scenario,''
  Phys.\ Rev.\ D {\bf 66}, 046005 (2002)
  [arXiv:hep-th/0109050];
   J.~Khoury, B.~A.~Ovrut, P.~J.~Steinhardt and N.~Turok,
  ``A brief comment on 'The pyrotechnic universe',''
  arXiv:hep-th/0105212;

\bibitem{tolley}
  A.~J.~Tolley, N.~Turok and P.~J.~Steinhardt,
  ``Cosmological perturbations in a big crunch / big bang space-time,''
  Phys.\ Rev.\ D {\bf 69}, 106005 (2004)
  [arXiv:hep-th/0306109].
 
\bibitem{cyclic}
  P.~J.~Steinhardt and N.~Turok,
  ``Cosmic evolution in a cyclic universe,''
  Phys.\ Rev.\  D {\bf 65}, 126003 (2002)
  [arXiv:hep-th/0111098].
 
\bibitem{other5d}
  T.~J.~Battefeld, S.~P.~Patil and R.~H.~Brandenberger,
  ``On the transfer of metric fluctuations when extra dimensions bounce or
  stabilize,''
  Phys.\ Rev.\ D {\bf 73}, 086002 (2006)
  [arXiv:hep-th/0509043].

\bibitem{lythandfriends}
  D.~H.~Lyth,
  ``The primordial curvature perturbation in the ekpyrotic universe,''
  Phys.\ Lett.\  B {\bf 524}, 1 (2002)
  [arXiv:hep-ph/0106153]; `The failure of cosmological perturbation theory in the new ekpyrotic
  scenario,''
  Phys.\ Lett.\  B {\bf 526}, 173 (2002)
  [arXiv:hep-ph/0110007];  R.~Durrer,
  ``Clarifying perturbations in the ekpyrotic universe: A web-note,''
  arXiv:hep-th/0112026;   R.~Brandenberger and F.~Finelli,
  ``On the spectrum of fluctuations in an effective field theory of the
  ekpyrotic universe,''
  JHEP {\bf 0111}, 056 (2001)
  [arXiv:hep-th/0109004];   J.~Martin, P.~Peter, N.~Pinto Neto and D.~J.~Schwarz,
  ``Passing through the bounce in the ekpyrotic models,''
  Phys.\ Rev.\  D {\bf 65}, 123513 (2002)
  [arXiv:hep-th/0112128];  J.~c.~Hwang,
  ``Cosmological structure problem in the ekpyrotic scenario,''
  Phys.\ Rev.\  D {\bf 65}, 063514 (2002)
  [arXiv:astro-ph/0109045].

\bibitem{paolonic}
  P.~Creminelli, A.~Nicolis and M.~Zaldarriaga,
  ``Perturbations in bouncing cosmologies: Dynamical attractor vs scale
  invariance,''
  Phys.\ Rev.\ D {\bf 71}, 063505 (2005)
  [arXiv:hep-th/0411270].

\bibitem{robertrecent}
  S.~Alexander, T.~Biswas and R.~H.~Brandenberger,
  ``On the Transfer of Adiabatic Fluctuations through a Nonsingular
  Cosmological Bounce,''
  arXiv:0707.4679 [hep-th].

\bibitem{pbbgw}
  R.~Brustein, M.~Gasperini, M.~Giovannini, V.~F.~Mukhanov and G.~Veneziano,
  ``Metric perturbations in dilaton driven inflation,''
  Phys.\ Rev.\  D {\bf 51}, 6744 (1995)
  [arXiv:hep-th/9501066];
   R.~Brustein, M.~Gasperini, M.~Giovannini and G.~Veneziano,
  ``Relic gravitational waves from string cosmology,''
  Phys.\ Lett.\  B {\bf 361}, 45 (1995)
  [arXiv:hep-th/9507017];
    M.~Gasperini,
  ``Tensor perturbations in high-curvature string backgrounds,''
  Phys.\ Rev.\  D {\bf 56}, 4815 (1997)
  [arXiv:gr-qc/9704045].

\bibitem{gwaves}
  L.~A.~Boyle, P.~J.~Steinhardt and N.~Turok,
  ``The cosmic gravitational wave background in a cyclic universe,''
  Phys.\ Rev.\  D {\bf 69}, 127302 (2004)
  [arXiv:hep-th/0307170].

\bibitem{chris}
  F.~Gmeiner and C.~D.~White,
  ``DBI Inflation using a One-Parameter Family of Throat Geometries,''
  JCAP {\bf 0802}, 012 (2008)
  [arXiv:0710.2009 [hep-th]].


\bibitem{seery}
  D.~Seery and J.~E.~Lidsey,
  ``Primordial non-gaussianities in single field inflation,''
  JCAP {\bf 0506}, 003 (2005)
  [arXiv:astro-ph/0503692].


\bibitem{daniel}
D.~Baumann, L.~Senatore, M.~Zaldarriaga,
  ``Scale-Invariance and the Strong Coupling Problem,''
    [arXiv:1101.3320 [hep-th]].


\bibitem{effinf}
  C.~Cheung, P.~Creminelli, A.~L.~Fitzpatrick, J.~Kaplan and L.~Senatore,
  ``The Effective Field Theory of Inflation,''
  JHEP {\bf 0803}, 014 (2008)
  [arXiv:0709.0293 [hep-th]].




\end{thebibliography}
\end{document}